%% file: ASE-main.tex
\documentclass[conference]{IEEEtran}
\IEEEoverridecommandlockouts
\usepackage{amsmath}
\usepackage{amsfonts}
\usepackage{algorithm}
\usepackage{algorithmic}
\usepackage{graphicx}
\usepackage{textcomp}
\usepackage{xcolor}
\usepackage{subfiles}
\usepackage{multirow}
\usepackage{booktabs}
\usepackage{vcell}
\usepackage{enumitem}
\usepackage{xspace}
\usepackage{hyperref}
\usepackage{colortbl}
\usepackage{tikz}
\usepackage{listings, color}
\usepackage{makecell}
\usepackage{array}
\usepackage{simplebnf}
\usepackage{subfig}
\usepackage{xstring} % 需要添加这个包来支持字符串操作
\makeatletter % changes the catcode of @ to 11
\newcommand{\linebreakand}{%
  \end{@IEEEauthorhalign}
  \hfill\mbox{}\par
  \mbox{}\hfill\begin{@IEEEauthorhalign}
}
\makeatother
\newcommand{\highlight}[2]{%
    \textcolor{black}{#2}%
}
\newcommand{\tool}{TRACE\xspace}
\newcommand{\coedpilot}{CoEdPilot\xspace}

\AtBeginDocument{%
  }
\begin{document}

\title{Learning Project-wise Subsequent Code Edits via Interleaving Neural-based Induction and Tool-based Deduction
\thanks{This research is supported in part by National Natural Science Fundation of China (62572300), the Minister of Education, Singapore (MOE-T2EP20124-0017, MOET32020-0004), the National Research Foundation, Singapore and the Cyber Security Agency under its National Cybersecurity R\&D Programme (NCRP25-P04-TAICeN), DSO National Laboratories under the AI Singapore Programme (AISG Award No: AISG2-GC-2023-008-1B), and Cyber Security Agency of Singapore under its National Cybersecurity R\&D Programme and CyberSG R\&D Cyber Research Programme Office. Any opinions, findings and conclusions or recommendations expressed in this material are those of the author(s) and do not reflect the views of National Research Foundation, Singapore, Cyber Security Agency of Singapore as well as CyberSG R\&D Programme Office, Singapore.}
}

% \author{
% \IEEEauthorblockN{Chenyan Liu}
% \IEEEauthorblockA{\textit{Shanghai Jiao Tong University}\\
% Shanghai, China \\
% \textit{National University of Singapore}\\
% Singapore, Singapore \\
% chenyan@u.nus.edu}
% \and
% \IEEEauthorblockN{Yun Lin\IEEEauthorrefmark{1}\thanks{\IEEEauthorrefmark{1} Corresponding author}}
% \IEEEauthorblockA{\textit{Shanghai Jiao Tong University}\\
% Shanghai, China\\
% lin\_yun@sjtu.edu.cn}
% \and
% \IEEEauthorblockN{Yuhuan Huang}
% \IEEEauthorblockA{\textit{Shanghai Jiao Tong University}\\
% Shanghai, China \\
% hyh0u0@sjtu.edu.cn}
% \linebreakand
% \IEEEauthorblockN{Jiaxin Chang}
% \IEEEauthorblockA{\textit{Shanghai Jiao Tong University}\\
% Shanghai, China \\
% cjx001234@sjtu.edu.cn}
% \and
% \IEEEauthorblockN{Binhang Qi}
% \IEEEauthorblockA{\textit{National University of Singapore}\\
% Singapore, Singapore \\
% qibh@nus.edu.sg}
% \and
% \IEEEauthorblockN{Bo Jiang}
% \IEEEauthorblockA{\textit{Bytedance Network Technology} \\
% Beijing, China \\
% jiangbo.jacob@bytedance.com}
% \linebreakand
% \IEEEauthorblockN{Zhiyong Huang}
% \IEEEauthorblockA{\textit{National University of Singapore}\\
% Singapore, Singapore \\
% huangzy@comp.nus.edu.sg}
% \and
% \IEEEauthorblockN{Jin Song Dong}
% \IEEEauthorblockA{\textit{National University of Singapore}\\
% Singapore, Singapore \\
% dcsdjs@nus.edu.sg}
% }

\author{
\IEEEauthorblockN{
Chenyan Liu\textsuperscript{1,}\textsuperscript{2},
Yun Lin\textsuperscript{1}\IEEEauthorrefmark{1},
Yuhuan Huang\textsuperscript{1},
Jiaxin Chang\textsuperscript{1},
Binhang Qi\textsuperscript{2},\\
Bo Jiang\textsuperscript{3},
Zhiyong Huang\textsuperscript{2},
and Jin Song Dong\textsuperscript{2}
}

\IEEEauthorblockA{\textsuperscript{1}Shanghai Jiao Tong University, Shanghai, China}
\IEEEauthorblockA{\textsuperscript{2}National University of Singapore, Singapore}
\IEEEauthorblockA{\textsuperscript{3}Bytedance Network Technology, Beijing, China\\
chenyan@u.nus.edu, \{lin\_yun, hyh0u0, cjx001234\}@sjtu.edu.cn, qibh@nus.edu.sg, \\ jiangbo.jacob@bytedance.com, huangzy@comp.nus.edu.sg, dcsdjs@nus.edu.sg}

\thanks{\IEEEauthorrefmark{1} Corresponding author}}

\maketitle

\begin{abstract}
% Recent years have seen the emergence of LLM-based solutions for software development.
% Practical software development can involve
% many incremental code edits over the project.
% Existing LLM-based code editing approaches rely on neural models to locate and generate subsequent edits.
% Despite spanning over the whole software project,
% practical code edits in an editing session may coherently compose a unified edit action.
% Such coherence is non-trivial and cannot be effectively captured by neural models relying on exhaustive search across the project,
% raising large challenges for the existing LLM-based state-of-the-arts
% regarding both the accuracy and the runtime overhead of edit location and generation.

% Recent years have seen the rise of AI-empowered automatic programming capability.
% Industrial and open-source developers usually need to accomplish project-wise code editing tasks such as 
% feature enhancement, refactoring, and bug fixing.
% Hence, predicting project-wise subsequent code editing can significantly improve their productivity.
% Researchers and industrial practitioners have
% proposed many solutions to automate the task.

In industrial and open-source software engineering tasks,
developers often perform project-wise code editing tasks, including 
feature enhancement, refactoring, and bug fixing,
where the leading AI models are expected to support the productivity.
Hence, researchers and practitioners have proposed and adopted many LLM-based solutions to facilitate their real-world development.
However, they largely suffer from the balance among
predicting scope, accuracy, and efficiency.
For example, solutions like Cursor achieve high accuracy only in a local editing scope 
while its performance drops on cross-file edits.
In contrast, solutions like CoEdPilot exhibit efficiency limitations when used to predict project-wise edits.

In this work, we propose \tool (\textbf{T}ool-integrated \textbf{R}ecom-mend\allowbreak\textbf{A}\allowbreak tion for \textbf{C}ode \textbf{E}diting),
a novel subsequent code editing solution to push the boundary of 
scope, accuracy, and efficiency.
Our rationale lies in that code edits are triggered for either semantic or syntactic reasons.
Therefore, \tool predicts subsequent edits by 
interleaving neural-based induction for semantic edit prediction and
tool-based deduction for syntactic edit prediction.
The tools can be any IDE facilities, such as refactoring tools (e.g., rename) or linting tools (e.g., use-def),
providing decent performance of \textit{deducing} edit-location and edit-generation.
Technically, we address the challenge of
(1) when to interleave between neural-based and tool-based prediction and
(2) how to further improve the performance of neural-based prediction.
As for the former, we learn a neural model to detect when to invoke IDE editing tools.
As for the latter, we propose a novel and fine-grained editing representation to further boost the performance of neural editing models.

Our extensive experiments show that,
in comparison to the state-of-the-arts such as CoEdPilot, GrACE, and CCT5,
\tool significantly improves the performance of edit location (by 43.76\%) and % (69.61/48.42-1)
edit generation (by 11.16\%). % (48.02/43.20-1)
Our simulation experiment on an interactive editing setting
shows that
\tool achieves an acceptance rate 6.15\% higher than Cursor.
Moreover, our user study consists of 24 participants on Cursor, CoEdPilot, and \tool, on three code editing tasks.
The results show that the experimental group with \tool achieves leading performance on cross-file global edits.
In addition, we observe concerning user behaviours on how participants deal with false predictions by the tools, shedding light on the design of future code-editing tools.
%has superior capability in project-wise subsequent editing compared with \coedpilot and Cursor,
%indicating \tool as a new state-of-the-art code editing solution
%in practice.
%The tool demonstration and experimental details are available at \url{https://sites.google.com/view/code-trace}.
\end{abstract}


\input{Sections/Intro}

\input{Sections/problem-statement}

\input{Sections/Motivation_new}

\input{Sections/Methodology}

\input{Sections/Experiment}

\input{Sections/user_study}

\input{Sections/Related_work}

\input{Sections/Conclusion}

\bibliographystyle{ieeetr} %ieeetr国际电气电子工程师协会期刊
\bibliography{sample-base}
% \vspace{12pt}
% \color{red}
% IEEE conference templates contain guidance text for composing and formatting conference papers. Please ensure that all template text is removed from your conference paper prior to submission to the conference. Failure to remove the template text from your paper may result in your paper not being published.
\end{document}

%% file: Sections/Intro.tex
\section{Introduction}
Recent years have witnessed a surge in applying large language models (LLM) for code generation \cite{feng2020codebert, guo2020graphcodebert, wang2021codet5, copilot, openai}.
Despite their success in translating natural language descriptions into target code,
incremental code edits across the project are more common in practice.
Empirical observation
shows that such incremental code edits account for over 70\% of changes over the code commit history \cite{nguyen2013study}.
It highlights the need for predicting project-wise subsequent code edits, 
which can significantly improve the productivity of software developers.
Existing works in the community have made the following efforts to address the problem.

\noindent\textbf{Local Edit Solution.}
A line of code edit works \cite{vaswani2017attention, grace, lin2023cct5, ZhangETAL22CoditT5, MODIT}
simplifies the edit generation problem into a machine translation task.
In general, these approaches
take as input the pre-edit code,
and generate the post-edit code as output.
Since those approaches cannot predict the edit location,
they are generally limited when being applied to predict project-wise subsequent edits.

\noindent\textbf{Project-wise Edit Solution.}
In this context,
several industry solutions, such as Copilot Edits and Cursor’s next edit suggestion \cite{cursor, copilot-edit}, have emerged.
However, due to latency and cost considerations,
these solutions appear to be conservative when recommending cross-file edits, often degrading into merely local edit suggestions.
When cross-file edits are expected,
users typically rely on features like Copilot Chat or Cursor Chat,
where they must specify files to edit, requiring prior knowledge and risking overlooked files.

Meanwhile, in the research community,
the most relevant work is \coedpilot \cite{code-edit-pilot},
which provides a project-wise solution regarding both edit location and generation,
by orchestrating a set of models.
A \textit{locator} model is trained to predict the edit type of each line of code in the code window.
If a line is predicted to be modified (e.g., insert or replace),
then a \textit{generator} model generates its edit solution.
Both the locator and generator models also incorporate prior edits as user feedback to infer the implicit edit specification.

Despite these advances toward a practical AI pair programmer solution,
they still suffer from the following challenges:

\begin{itemize}[noitemsep, topsep=0pt, leftmargin=*]
  \item \textbf{C1 (Location Overhead): }
    A software project can be large,
    meaning the model may process numerous code windows to locate subsequent edits.
    Therefore, exhaustively monitoring the whole project upon any code edits,
    %which is triggered upon any code edits, is computationally expensive,
    can incur non-trivial runtime overhead, and undermine the performance. % of edit prediction.
  \item \textbf{C2 (Overlooked Edit Composition): }\highlight{C3, C8}{
    Existing neural-based solutions predict each code edit as an individual \textit{hunk} (i.e., consecutive lines of code change, see example hunks in \autoref{tab:example1}).
    Despite its rapid evolution, neural-based induction inherently suffers from non-trivial computational overhead for scanning the entire codebase and remains fundamentally prone to hallucinations \cite{xu2024hallucination}.
    Moreover, \cite{su2025code} shows that LLMs perform poorly on static analysis tasks and that pre-training on such tasks does not improve general code intelligence.
    However, in practice, hunks can be highly associated or occur simultaneously,
    forming high-level actions like refactoring.
    We refer to these grouped edits as \textbf{edit compositions},
    where edits exhibit \textbf{coherence}, meaning they are likely to propagate to each other, and this propagation can often be captured by static analysis tools, which guarantee both speed and correctness.} 
  \item \textbf{C3 (Coarse-grained Edit Representation): }
    The existing state-of-the-arts \cite{vaswani2017attention, grace, lin2023cct5, ZhangETAL22CoditT5, MODIT} adopt the git-diff-style representation \cite{gitdiff}, modeling code edits as replacements or insertions (see \autoref{tab:example2} H1 as an example).
    While being widely adopted,
    such a coarse-grained edit representation
    can express semantically different editing scenarios in a similar way (see \autoref{tab:example2} H2 as an example),
    leading to training inefficiency when learning the code-editing models.
\end{itemize}

\highlight{C1}{
To address the above challenges, 
we propose \tool (\textbf{T}ool-integrated \textbf{R}ecommend\textbf{A}tion for \textbf{C}ode \textbf{E}diting), 
predicting project-wise subsequent code edits primarily based on a set of prior edits. 
Our rationale lies in that code edits are triggered for either semantic or syntactic reasons. 
Therefore, \tool predicts subsequent edits by interleaving 
\textbf{neural-based induction} for semantic edit prediction and 
\textbf{tool-based deduction} for syntactic edit prediction.
Technically, \tool recommends subsequent edits through a pipelined workflow. 
At each step, an \textit{edit-composition invoker} first monitors the ongoing session 
and checks whether the prior edits match pre-defined edit compositions. 
In such cases, TRACE applies \textbf{tool-based deduction}, proactively invoking IDE services (e.g., rename, use-def update, or remove unused imports). 
This mechanism leverages the syntactic coherence among edits in a composition to quickly narrow down candidate locations, 
especially when edits span multiple files. 
If no tool service is triggered, \tool falls back to \textbf{neural-based induction}, 
which applies a sliding window over the project with
our \textit{edit locator} and \textit{edit generator}.
Both are equipped with the novel edit representation to distinguish diverse edit scenarios (see \autoref{tab:example2}).
In this way, TRACE integrates tool invocation with neural inference into a feedback-driven pipeline that progressively recommends coherent edits.}
% --------------- after -----------------

% % --------------- before -----------------
% % Summarize experiment
% We extensively evaluate \tool on 38K code commits,
% spanning over 678 source projects of 5 programming languages.
% The results show that
% (1) in comparison to the state-of-the-arts,
% \tool can significantly improve the precision of edit location by 43.76\%,
% the recall of that by 9.96\%,
% and the exact match of edit generation by 11.16\%,
% (2) \tool can well identify edit composition and invoke the tool service at the appropriate time (92.45\% precision and 94.63\% recall),
% (3) our proposed edit representation can also significantly boost the performance of the neural edit locator by 14.57\% and the generator by 7.40\%.
% In addition, our edit simulation experiment,
% \tool reduces the time cost by 14.40\% and achieves edit suggestion acceptance of 27.71\%, on par with Cursor.
% Last but not least, our user study of 24 participants on 3 editing tasks shows that \tool is capable of effectively recommending project-wise subsequent edits, especially on those cross-file edits.
% % --------------- before -----------------

% --------------- after -----------------
We extensively evaluate tool on 38K code commits from 678 projects across 5 programming languages. 
Compared to state-of-the-arts, 
(1) \tool significantly improves edit location precision by 43.76\%, recall by 9.96\%, 
and edit generation accuracy by 11.16\%. 
(2) \tool can well identify edit composition and invoke tools appropriately (92.45\% precision and 94.63\% recall),
(3) The novel edit representation enhances the neural edit locator by 14.57\% and generator by 7.40\%. 
In edit simulation, tool reduces time cost by 14.40\% and achieves 27.71\% suggestion acceptance, comparable to Cursor. 
A user study with 24 participants on 3 tasks confirms TRACE’s effectiveness in recommending project-wise, cross-file edits.
% --------------- after -----------------

% Summairze contribution
Overall, we summarize our contributions as follows:
\begin{itemize}[noitemsep, topsep=0pt, leftmargin=*]
    \item \textbf{Methodology.}
        To the best of our knowledge,
        we are the first to propose 
        to predict project-wise subsequent edits by interleaving neural-based induction and tool-based deduction.
        This solution can largely mitigate the LLM hallucination and improve the runtime efficiency (especially for edit location).
        %learning the invocation of the static analysis tool services for more accurate project-wise edit prediction.
        In addition, we discover a more expressive edit representation to further advance the performance of the neural edit locator and generator.
    \item \textbf{Tool.}
        We implement \tool as a Visual Studio Code (VS Code) extension \cite{vscode_extension} for interactive edit localization and generation.
        \tool is designed to smoothly integrate Language Service Protocol (LSP) invocation and neural edit location/generation,
        potentially helping programmers accomplish their tasks in practice.
    \item \textbf{Evaluation.}
        We extensively and systematically evaluate \tool on 38K code commits, spanning over 678 source projects and 5 programming languages,
        via benchmarks and real-world editing simulation,
        establishing \tool as the new state-of-the-art project-wise code editing solution.
    \item \textbf{User Study.}
        We further conducted a user study with 24 participants over 3 editing tools, totalling about 118 man-hours.
        The results confirm the effectiveness of our design in practice,
        also revealing a concerning \textit{over-trust} phenomenon among participants,
        regardless of their experience.
        This highlights directions for improving future code-editing solutions.
\end{itemize}

\highlight{C4}{Additional demonstration videos, source code, supplementary experimental information, and user study video recordings are available at our homepage \cite{homepage}}.
% , \highlight{C3, C13}{also archived at \cite{archivehomepage}}.

%% file: Sections/problem-statement.tex
\section{Problem Statement} 
\label{sec:problem-statement}
In this work, the problem statement is formulated as follows. Given the following inputs:
\begin{itemize}[leftmargin=*]
    \item A software project $P = \{f_1, \ldots, f_n\}$, where $f_i$ is a source file in the project,
    \item A sequence of prior edits as $E_p=\langle e_1, \ldots,e_k\rangle$ in an edit session, where the subscript $i\in[0, k]$ is the chronological order. Each edit $e\in E_p$ is defined as $e = (f, line_{start}, line_{end}, code_b, code_a)$, where $f\in P$ denotes the file where the edit $e$ happens, $line_{start}$ and $line_{end}$ denote the scope of $e$ in $f$, $code_b$ and $code_a$ denote code before and after the edit in the scope $(line_{start}, line_{end})$.
    \item An optional edit description $prompt$.
\end{itemize}
our \tool solution is to generate $e_{k+1}$ in the editing session based on $P$, $E_{p}$, and $prompt$ (optional).

Different from the solutions \cite{yang2024swe, zhang2024autocoderover, zhang2024codeagent} of issue-resolving tasks like SWE-bench \cite{jimenez2023swe} 
which results in a set of patches in a project aligning with an issue description,
our task is more progress-driven.
Specifically, we predict a subsequent edit aligning with the flow of prior edits with an optional prompt.
Note that not all the code editing tasks are driven by a detailed issue description.

%% file: Sections/Motivation_new.tex
\section{Motivating Example} \label{sec:example}

\subsection{Tool-inducing Edit Composition}
\autoref{tab:example1} shows four \textit{hunks} (denoted by H$_i$) over two source files, extracted from commit\footnote{https://github.com/pingcap/tidb/commit/fcef061} in project \texttt{ping/tidb}. 
Hunks can be summarized into two actions (A1 and A2):
\begin{itemize}[noitemsep, topsep=0pt, leftmargin=*]
  \item \textbf{A1: Method Signature Update:}
    The function \texttt{renew\allowbreak With\allowbreak Capacity()} is updated to include a new parameter \texttt{max\allowbreak Chunk\allowbreak Size},
    which includes H1 (line 1-4), H3 (line 9-10), and H4 (line 11-12).
    The three hunks are mutually depen-dent, as the method signature update propagates across them.
  \item \textbf{A2: Attribute Initialization Update:}
    H2 (lines 5–8) updates the initialization of \texttt{newChk.requiredRows} from variable \texttt{cap} to \texttt{maxChunkSize}.
\end{itemize}
\input{Sections/example1}

\vspace{-10pt}
The state-of-the-art edit locator, \coedpilot \cite{code-edit-pilot}, predicts these hunks sequentially.
As for the first action (A1),
given a prior edit as H1,
it requires an LLM to analyze the \textit{entire} project code to infer whether and what changes will be propagated.
The challenge lies in
the substantial computational overhead required to scan the entire codebase,
as well as the risk of false positives in similar textual patterns,
such as \texttt{newWithCapacity()} and \texttt{renewColumns\allowbreak WithCapacity()},
caused by LM hallucination. 
% To circumvent these challenges,
% Cursor adopts a conservative strategy that confines next edit prediction to localized code regions,
% while \coedpilot introduces a file locator mechanism to perform a coarse-grained selection of relevant files,
% thereby narrowing the search space.

In this work, we mitigate such overhead and false positives by learning to invoke
the inherent static tools in IDE.
Specifically, we monitor the code-editing session and infer when to invoke a static tool based on prior edits (e.g., H1), 
to retrieve the rest of edit composition (e.g., H3 and H4),
which can
(1) significantly lower the cost and
(2) boost the performance of
edit location and generation.
% In this example,
% in contrast to \coedpilot, which predicts the four hunks (H1-4) one after another,
% which leads to repeated inference and project-wide scanning,
% \tool can infer the location and edit content of an edit composition (H1, H3, and H4) with higher precision and efficiency.

\begin{figure}[!htb]
%\vspace{-5pt}
\centering
% 左图
\begin{minipage}[t]{0.45\linewidth}
    \centering
    \includegraphics[width=\linewidth]{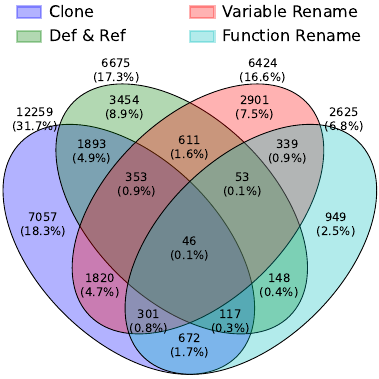}
    \caption{Percentage of commits with edit composition.}
    \label{fig:empirical_composition}
\end{minipage}%
\hfill
% 右表，顶部补空使其与图顶对齐
\begin{minipage}[t]{0.5\linewidth}
    \vspace{-2.6cm}  % 可微调
    \centering
    \footnotesize
    \tabcolsep=0.1cm
    \begin{tabular}{c|c}
    \hline
    \textbf{Language} & \makecell{\textbf{Percentage} \\ \textbf{(\%)}} \\
    \hline
    Python & 17.76 \\
    Go & 18.32 \\
    Java & 17.03 \\
    JavaScript & 20.81 \\
    TypeScript & 19.20 \\
    \hline
    \textbf{Avg.} & 18.04 \\
    \hline
    \end{tabular}
    \captionsetup{type=table}
    \caption{Percentage of edit hunks with multiple semantics.}
    \label{tab:empirical_representation}
\end{minipage}
\vspace{-5pt}
\end{figure}

\highlight{C2}{
\textbf{Empirical Study.} To validate the wide applicability of our approach, we empirically measure the frequency of edit compositions in real-world commits.
We collected a dataset of 38K commits from top-starred GitHub repositories across 5 languages (dataset detail refers to \autoref{tab:benchmark} in Section~\ref{sec:experiment} Experiment Setup).
Following the benchmark construction methodology described in Section~\ref{subsec:rq4}, 
we applied LSP-based analysis to detect four types of edit compositions: variable renames, function renames, definition/reference lookups, and code clones.
\autoref{fig:empirical_composition} shows that 53.6\% of commits contain at least one type of predefined edit composition detectable by LSP services, 
with 16.4\% involving multiple types,
indicating a high frequency of edit composition in real-world editing scenarios.}

\textbf{Challenges.}
To deliver this potential, \tool needs to address \textit{when} and \textit{what} edit composition to invoke.
First, an IDE may support various edit compositions,
which may result in misleading invocation opportunities.
For example, H2 in \autoref{tab:example1} may indicate a \textit{renaming} edit, as it replaces all instances of \texttt{cap} with \texttt{maxChunkSize}.
However, this reflects a usage change rather than a true renaming edit. 
Applying a rename tool here may introduce incorrect changes across multiple locations and files, which can be costly to undo the changes.
Second, even with accurate composition prediction, 
key details may remain missing.
For example, knowing H1 triggers a \textit{signature update} still leaves the parameter choices in H3 and H4 to be inferred.

Thus, we use \tool as a hybrid solution combining an edit-composition invoker,
a neural edit locator and a neural edit generator
that complement each other.
On the one hand, the edit-composition invoker can speed up the edit locator and edit generator with higher confidence.
On the other hand, the edit locator and edit generator can complement the missing details for the edit-composition invoker.
This enables \tool to better predict project-wise code edits.

\subsection{Edit Representation}
\autoref{tab:example2} shows a commit\footnote{https://github.com/AUTOMATIC1111/stable-diffusion-webui/commit/9e27af7} from \texttt{stable-\allowbreak diffusion-\allowbreak webui}, 
A web interface for Stable Diffusion models,
motivating us to further design a new edit representation.
Existing works \cite{code-edit-pilot, lin2023cct5, grace}
represent code edits as code lines labelled with \textit{replace}, \textit{insert}, and \textit{keep}.
For example, for H1 in \autoref{tab:example2},
line 1 has a tag of \textless KEEP\textgreater\ and line 2 has a tag of \textless REPLACE\textgreater.
Existing solutions follow the git-diff representation as shown in H1, H2, and H3.

However, we observe that such a representation represents different scenarios in a similar way,
which can cause training confusion.
For example, the representation in H1 indicates that the \texttt{if} condition (line 2) should be changed into
a code block (line 3-5 in H1),
which can hardly be learned to generalized to H2 and H3.
Nevertheless, probing into the details of H1 indicates that
the \textbf{single} hunk contains \textbf{two} edit semantics: both \textit{insert} (line 2,3 in H1') and \textit{replace} (line 4,5 in H1'), which are under-represented by the unified \textit{replace} label in git diff.
In other words, reformulating the edit representation as H1' enables
the model to generalize the replacement at line 5 (in H1') to H2 and H3
in a far more convenient way during model training.

\highlight{C2}{
\textbf{Empirical Study.}
To validate the prevalence of such multi-semantic hunks, 
we apply \autoref{alg:representation} in Section~\ref{sec:edit-representation} to analyze the number of distinct edit semantics contained in each edit hunk from our dataset.
We compute the percentage of edit hunks that encompass more than one edit semantic (e.g., code deletion and replacement within the same hunk).
\autoref{tab:empirical_representation} shows that 18.04\% of edit hunks involve multiple edit semantics that are not fully captured by the traditional git-diff format.}

Given the observation, we propose a new edit representation with 6 editing labels
in contrast to 3 traditional editing labels (i.e., \textit{replace}, \textit{insert}, and \textit{keep}),
further boosting the performance of neural edit locator and generator. \input{Sections/example2}

\vspace{-10pt}

%% file: Sections/example1.tex
\definecolor{codegreen}{rgb}{0.0, 0.66, 0.30}
\definecolor{codered}{rgb}{1,0.05,0.05}
\definecolor{codegray}{rgb}{0.5,0.5,0.5}
\definecolor{codegray}{rgb}{0.5,0.5,0.5}
\definecolor{codeblue}{rgb}{0.0,0.0,0.5}
\definecolor{codered}{rgb}{0.6,0.0,0.0}
\definecolor{codepurple}{rgb}{0.58,0.0,0.82}
\definecolor{codeorange}{rgb}{1.0,0.5,0.0}
\definecolor{codecyan}{rgb}{0.0,0.6,0.6} % New color for strings

\lstdefinestyle{lgeometry}{ xleftmargin = 20pt, xrightmargin = 0pt, frame = tb, framesep = \fboxsep, framexleftmargin = 20pt}

% Define a style for the listings
\lstdefinestyle{gostyle}{
    language=Go,
    basicstyle=\ttfamily\tiny,
    breakatwhitespace=false,
    breaklines=true,
    captionpos=b,
    keepspaces=true,
    showspaces=false,
    showstringspaces=false,
    showtabs=false,
    tabsize=2,
    commentstyle=\color{codegray},
    keywordstyle=\color{blue},
    stringstyle=\color{codepurple},
    moredelim=[is][\color{codered}\textbf]{@@}{@@},
    moredelim=[is][\color{codegreen}\textbf]{^}{^},
    columns=flexible,
    aboveskip=-0.5\baselineskip,
    belowskip=-1.6\baselineskip,
    numbers=left,
    numberstyle=\tiny\color{codegray},
    stepnumber=1,
    firstnumber=auto,
    numbersep=4pt,
    % style = lgeometry,
}
\lstset{style=gostyle}

%\clearpage
\noindent
\begin{table}[t]
\caption{
Example 1: Composite edit v.s. individual edit
}
%\vspace{-5pt}
\label{tab:example1}
%\scriptsize
\centering
\small
\tabcolsep=0.06cm
\begin{tabular}{|m{0.04\linewidth}|p{0.02\linewidth}|p{0.90\linewidth}|}
%\hline
% https://github.com/pingcap/tidb/commit/fcef0610597763d1dcdb8b86faa0b4312abc881b
%\makecell[l]{\textbf{Hunk}} & \multicolumn{2}{|c|}{\textbf{Code}}  \\
\hline
\multicolumn{3}{|c|}{\textbf{File:} executor/window.go} \\
\hline
\multirow{4}{*}{H1} & & \cellcolor{red!20}
\begin{lstlisting}[firstnumber=1]
- func renewWithCapacity(chk *Chunk,cap int) *Chunk {
\end{lstlisting} \\
& & \cellcolor{green!20}
\begin{lstlisting}[firstnumber=2]
+ func renewWithCapacity(chk *Chunk,cap,maxChunkSize int) *Chunk {
\end{lstlisting} \\
& &
\begin{lstlisting}[firstnumber=3]
      newChk := new(Chunk)
      ...
\end{lstlisting} \\
\hline
\multirow{4}{*}{H2}
& &
\begin{lstlisting}[firstnumber=5]
      newChk.capacity = cap
\end{lstlisting} \\
& & \cellcolor{red!20}
\begin{lstlisting}[firstnumber=6]
-     newChk.requiredRows = cap
\end{lstlisting} \\
& & \cellcolor{green!20}
\begin{lstlisting}[firstnumber=7]
+     newChk.requiredRows = maxChunkSize
\end{lstlisting} \\
& &
\begin{lstlisting}[firstnumber=8]
      return newChk
\end{lstlisting}
\\
\hline
\multirow{2}{*}{H3}
& & \cellcolor{red!20}
\begin{lstlisting}[firstnumber=9]
-     return renewWithCapacity(chk, newCap)
\end{lstlisting} \\
& & \cellcolor{green!20}
\begin{lstlisting}[firstnumber=10]
+     return renewWithCapacity(chk, newCap, maxChunkSize)
\end{lstlisting} \\
\hline
\multicolumn{3}{|c|}{\textbf{File:} util/chunk/row.go} \\
\hline
\multirow{2}{*}{H4}
& & \cellcolor{red!20}
\begin{lstlisting}[firstnumber=11]
-     newChk := renewWithCapacity(r.c, 1)
\end{lstlisting} \\
& & \cellcolor{green!20}
\begin{lstlisting}[firstnumber=12]
+     newChk := renewWithCapacity(r.c, 1, 1)
\end{lstlisting} \\
\hline

\end{tabular}
\vspace{-15pt}
\end{table}

%% file: Sections/example2.tex
\definecolor{codegreen}{rgb}{0.0, 0.66, 0.30}
\definecolor{codered}{rgb}{1,0.05,0.05}
\definecolor{codegray}{rgb}{0.5,0.5,0.5}
\definecolor{codegray}{rgb}{0.5,0.5,0.5}
\definecolor{codeblue}{rgb}{0.0,0.0,0.5}
\definecolor{codered}{rgb}{0.6,0.0,0.0}
\definecolor{codepurple}{rgb}{0.58,0.0,0.82}
\definecolor{codeorange}{rgb}{1.0,0.5,0.0}
\definecolor{codecyan}{rgb}{0.0,0.6,0.6} % New color for strings

\lstdefinestyle{lgeometry}{ xleftmargin = 20pt, xrightmargin = 0pt, frame = tb, framesep = \fboxsep, framexleftmargin = 20pt}

% Define a style for the listings
\lstdefinestyle{gostyle}{
    language=Go,
    basicstyle=\ttfamily\tiny,
    breakatwhitespace=false,
    breaklines=true,
    captionpos=b,
    keepspaces=true,
    showspaces=false,
    showstringspaces=false,
    showtabs=false,
    tabsize=2,
    commentstyle=\color{codegray},
    keywordstyle=\color{blue},
    stringstyle=\color{codepurple},
    moredelim=[is][\color{codered}\textbf]{@@}{@@},
    moredelim=[is][\color{codegreen}\textbf]{^}{^},
    columns=flexible,
    aboveskip=-0.5\baselineskip,
    belowskip=-1.6\baselineskip,
    numbers=left,
    numberstyle=\tiny\color{codegray},
    stepnumber=1,
    firstnumber=auto,
    numbersep=4pt,
    % style = lgeometry,
}
\lstset{style=gostyle}

\noindent
\begin{table}[t]
\caption{
Example 2: Edit representation beyond replacement
}
\label{tab:example2}
%\scriptsize
\centering
%\vspace{-5pt}
\small
\tabcolsep=0.06cm
\begin{tabular}{|m{0.05\linewidth}|p{0.02\linewidth}|p{0.84\linewidth}|}
%\hline
% https://github.com/pingcap/tidb/commit/fcef0610597763d1dcdb8b86faa0b4312abc881b
%\makecell[l]{\textbf{Hunk}} & \multicolumn{2}{|c|}{\textbf{Code}}  \\
\hline
\multicolumn{3}{|c|}{\textbf{File:} modules/sd\_samplers\_kdiffusion.py} \\
\hline
\multirow{6}{*}{H1} & &
\begin{lstlisting}[firstnumber=1]
  extra_params_kwargs = self.initialize(p)
\end{lstlisting}\\
 & & \cellcolor{red!20}
\begin{lstlisting}[firstnumber=2]
- if 'sigma_min' in inspect.signature(self.func).parameters:
\end{lstlisting} \\
& & \cellcolor{green!20}
\begin{lstlisting}[firstnumber=3]
+ parameters = inspect.signature(self.func).parameters
+ xi = x + noise * sigma_sched[0]
+ if 'sigma_min' in parameters:
\end{lstlisting} \\
& &
\begin{lstlisting}[firstnumber=6]
      extra_params_kwargs['sigma_min'] = sigma_sched[-2]
\end{lstlisting} \\
\hline
\multirow{6}{*}{H1'} & &
\begin{lstlisting}[firstnumber=1]
  extra_params_kwargs = self.initialize(p)
\end{lstlisting}\\
& & \cellcolor{green!20}
\begin{lstlisting}[firstnumber=2]
+ parameters = inspect.signature(self.func).parameters
+ xi = x + noise * sigma_sched[0]
\end{lstlisting} \\
 & & \cellcolor{red!20}
\begin{lstlisting}[firstnumber=4]
- if 'sigma_min' in inspect.signature(self.func).parameters:
\end{lstlisting} \\
& & \cellcolor{green!20}
\begin{lstlisting}[firstnumber=5]
+ if 'sigma_min' in parameters:
\end{lstlisting} \\
& &
\begin{lstlisting}[firstnumber=6]
      extra_params_kwargs['sigma_min'] = sigma_sched[-2]
\end{lstlisting} \\
\hline
\multirow{4}{*}{H2}
& &
\begin{lstlisting}[firstnumber=7]
      ...
\end{lstlisting} \\
& & \cellcolor{red!20}
\begin{lstlisting}[firstnumber=8]
- if 'n' in inspect.signature(self.func).parameters:
\end{lstlisting} \\
& & \cellcolor{green!20}
\begin{lstlisting}[firstnumber=9]
+ if 'n' in parameters:
\end{lstlisting} \\
& &
\begin{lstlisting}[firstnumber=10]
      extra_params_kwargs['n'] = len(sigma_sched) - 1
\end{lstlisting}
\\
\hline
\multirow{4}{*}{H3}
& &
\begin{lstlisting}[firstnumber=11]
      ...
\end{lstlisting} \\
& & \cellcolor{red!20}
\begin{lstlisting}[firstnumber=12]
- if 'sigma_sched' in inspect.signature(self.func).parameters:
\end{lstlisting} \\
& & \cellcolor{green!20}
\begin{lstlisting}[firstnumber=13]
+ if 'sigma_sched' in parameters:
\end{lstlisting} \\
& &
\begin{lstlisting}[firstnumber=14]
      extra_params_kwargs['sigma_sched'] = sigma_sched
\end{lstlisting}
\\
\hline

\end{tabular}
\vspace{-10pt}
\end{table}

%% file: Sections/Methodology.tex
\section{Methodology}
\begin{figure}
    \centering
    \includegraphics[width=\linewidth]{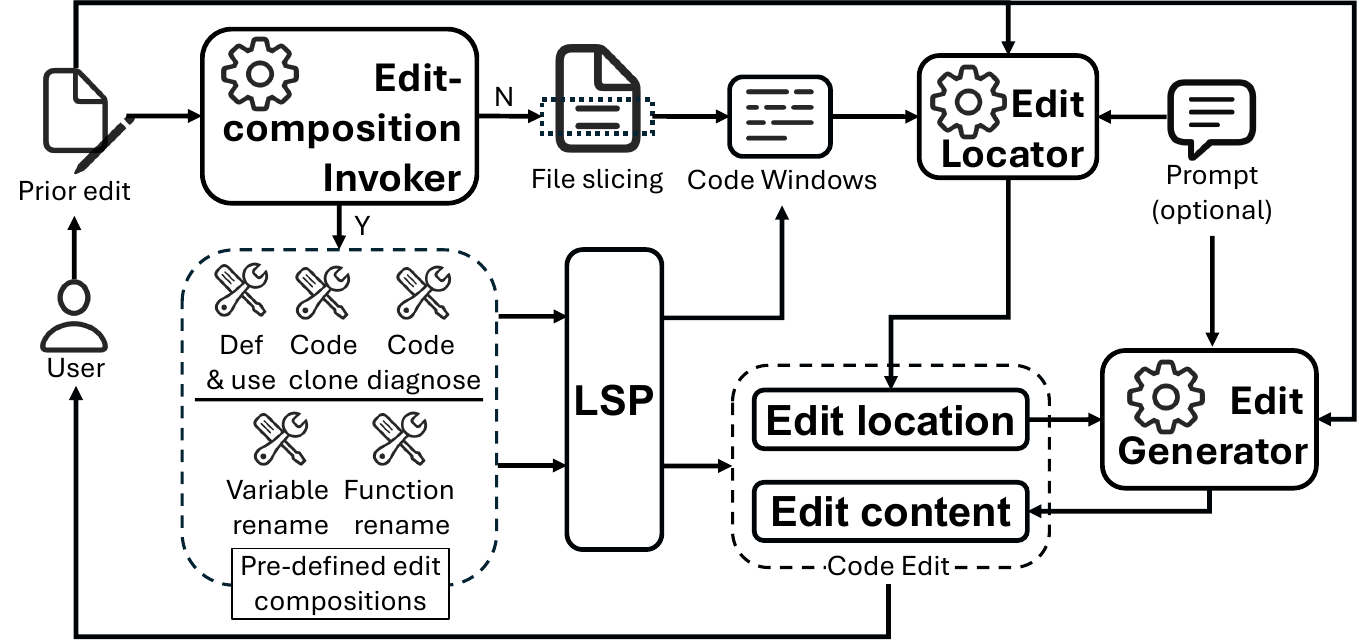}
    %\vspace{-5pt}
    \caption{Overview of \tool: 
    \tool generates code edits (in terms of the edit location and the edit content)
    by orchestrating three models, i.e., 
    edit composition invoker, edit locator, and edit generator.
    The generated code edits can serve as new prior edits to further
    generate new edits.
    }
    \label{fig:overview}
    %\vspace{-5pt}
\end{figure}

\highlight{C3, C12}{\autoref{fig:overview} shows an overview of \tool}, which takes the code project $P$, the user’s prior edits $E_{p}$, and optional edit description $prompt$ as input, and recommends the subsequent code edit $e_{k+1}$ in terms of both location and content.
The generated code edit serves as a new prior edit to predict the subsequent edits.
%making \tool an \highlight{AI pair programmer}, recommending edits iteratively and iteratively.
\tool orchestrates three neural network models,
which function as follows:\
\begin{itemize}[leftmargin=*]
  \item \textbf{Edit-composition invoker}:
    Given the prior edits $E_{p}$, the edit-composition invoker checks whether the last edit $e_k$ is part of a pre-defined edit composition (e.g., variable rename and def-use update) and, if so, invokes tool services to retrieve the remaining edits within the same composition.
  \item \textbf{Edit locator}:
    Given a code window as a scope, optional prompt $prompt$, and prior edits $E_{p}$, the edit locator predicts the edit label of each line of code and each space between lines.
    These labels serve as indicators for
    (1) the editing location within the scope and
    (2) the edit types at those locations, facilitating the follow-up edit content generation.
  \item \textbf{Edit generator}:
    Based on the predicted edit types in the code window, the optional prompt $prompt$, and $E_p$,
    the edit generator predicts the edit content $code_a$.
\end{itemize}

\highlight{C3, C6, C10}{
Once the latest edit $e_k$ is applied to the codebase, 
it is passed to the edit-composition invoker. 
If $e_k$ is identified as part of a predefined edit composition, the corresponding tool service will be automatically triggered.
Tools such as variable and function renaming from LSP provide both edit location and content, allowing us to skip the locator and generator steps.
Other tools may only provide approximate code windows, which require further processing by the edit locator and generator for contextual analysis.
The edit locator analyzes these windows to label each line and inter-line position based on context, determining precise editing lines and plans. 
Subsequently, the generator combines contextual information to produce the specific edit content for each identified location.
If no composition is detected, \tool slices the project into code windows and feeds them into the edit locator for labelling.
Only code windows that receive edit-requiring labels (e.g., \textbf{\textless REPLACE\textgreater}, \textbf{\textless INSERT\textgreater}) are subsequently passed to the generator for edit content generation.
}
In this work, we train code LLMs as the edit locator and edit generator.
Edit adopted by the user will be added to prior edits for the next recommendation.
\highlight{C4}{Given the space limit, model training hyper-parameters and input/output examples are available at \cite{homepage}}. 

\input{Sections/edit-representation.tex}

\input{Sections/edit-composition-invoker.tex}

\input{Sections/locator-and-generator.tex}

%% file: Sections/edit-representation.tex
\subsection{Edit Representation} \label{sec:edit-representation}

%\highlight{An edit representation encodes a sequence of action labels $op$ on the pre-edit code $code_b$, which specify how each part of the code should be transformed.}
Given an edit $e$, its edit representation specifies what edit operation type (e.g., insert, keep, replace, delete) is applied on each line of code in $code_b$.
Generally, an edit specified in Section~\ref{sec:problem-statement} can have many representations,
which makes an impact on the model training effectiveness.
In this work, we report a training-friendly representation for the model to distinguish different editing scenarios.
We used BNF to define our designed edit representation.
As shown in \autoref{fig:bnf}, 
%the comma stands for the concatenation of strings. 
an edit representation ($R_{edit}$) consists of 
two inter-line edit labels ($L_{inter}$) and a code representation ($R_{code}$).
A code representation consists of line representations ($R_{line}$), the edit label of a code line ($L_{inline}$), and a line code ($LoC$).
The edit labels are annotated with the edit type
defined as follows.
\begin{itemize}[noitemsep, topsep=0pt, leftmargin=*]
    \item \textbf{\textless KEEP\textgreater} ($L_{inline}$): No edit to conduct for a line;
    \item \textbf{\textless REPLACE\textgreater} ($L_{inline}$): A line of code is to modify;
    \item \textbf{\textless DELETE\textgreater} ($L_{inline}$): A line of code is to delete.
    \item \textbf{\textless NULL\textgreater} ($L_{inter}$): No code to insert between lines;
    \item \textbf{\textless INSERT\textgreater} ($L_{inter}$): New code is to insert between lines;
    \item \textbf{\textless BLOCK-SPLIT\textgreater} ($L_{inter}$): 
    A block consists of continuous lines with the same edit type.
    This edit label is to split two blocks with different semantics.
\end{itemize}

\begin{figure}[t]
  \centering
  
  \fbox{
    \begin{minipage}{0.95\linewidth}  % 使框与文本匹配
    \scriptsize
    \hspace*{-1em}  % 控制整体左移
    \begin{bnf}
    $R_{edit}$ : \textsf{Edit Representation} ::=
        $L_{inter}, R_{code}, L_{inter}$
        ;;
    $R_{code}$ : \textsf{Code Representation} ::= $R_{line}$ | $R_{line}, L_{inter}, R_{code}$
        ;;
    $R_{line}$ : \textsf{Line Representation} ::= $L_{inline}, LoC $
        ;;
    $LoC$ : \textsf{Line of Code} ::= one line of code
        ;;
    $L_{inter}$ : \textsf{Inter-line Label} ::= \textless NULL\textgreater | \textless INSERT\textgreater | \textless BLOCK-SPLIT\textgreater
        ;;
    $L_{inline}$ : \textsf{Inline Label} ::= \textless KEEP\textgreater | \textless REPLACE\textgreater | \textless DELETE\textgreater 
        ;;
    \end{bnf}
    \end{minipage}
  }
  %\vspace{-5pt}
  \caption{Edit representation of \tool in BNF, more expressive for different code editing scenarios, 
  with six edit labels.}\label{fig:bnf}
  \vspace{-10pt}
\end{figure}

% $L_{inter}$ describes how space between lines should change:
% \begin{itemize}[noitemsep, topsep=0pt, leftmargin=*]
%     \item \textbf{\textless NULL\textgreater} ($L_{inter}$): No code to insert;
%     \item \textbf{\textless INSERT\textgreater} ($L_{inter}$): New pieces of code should be inserted;
%     \item \textbf{\textless BLOCK-SPLIT\textgreater} ($L_{inter}$): The separation of two adjacent but differently edited \textless REPLACE\textgreater{} blocks, with no edit operation to conduct.
% \end{itemize}
\input{Sections/edit-representation-example-new.tex}

%\vspace{-10pt}
\noindent\textbf{An example}. 
\autoref{tab:representation-example} illustrates how a real edit hunk from project \texttt{localstack/localstack} \footnote{see https://github.com/localstack/localstack/commit/667c6c5.}
is translated into our representation.
The first column contains inter-line labels and the second contains inline labels.
Compared to the git-diff format,
the representation decomposes coarse-grained differences into finer-grained edit semantics.
While the git-diff format captures the entire hunk as a single replace,
\tool derives 6 edit semantics, i.e., a delete change (i.e., delete line 3); a replace change (i.e., replace line 4) and an insert change (i.e., insert between line 4, 5), etc.
This granularity enables more precise generalisation,
improving both edit localisation and generation (see the results in Section~\ref{subsec:rq1} and Section~\ref{subsec:rq2}).

\autoref{alg:representation} shows how an edit hunk from git-diff format is translated into our representation,
by aligning code lines before (e.g. old line 3-6 in \autoref{tab:representation-example}) and after the edit (e.g. new line 3-9).
Since no identical lines exist between the two versions, direct line-level matching is infeasible.
In \autoref{alg:representation} line 3, 
We use Tree-sitter \cite{treesitter} to tokenise both versions into syntax elements with their types, code text, and line indices. 
Longest Common Sub-sequence (LCS) is applied to the token sequences to match elements that remain unchanged, where a match is defined as having both identical type and code text. 
Based on the line index of matched elements, $token2line\_mapping(\cdot)$ builds line-level alignment.
Hence in \autoref{tab:representation-example}, old line 3 maps to $\emptyset$, old line 4 to new line 3 and $\emptyset$ to new lines 8, 9.
We then assign inline and inter-line labels according to each mapping block via $convert2label(\cdot)$:
matched lines in the old version are labelled with \textless REPLACE\textgreater;
lines in the old version that have no counterpart are labelled \textless DELETE\textgreater; 
for unmatched lines in the new version, the corresponding inter-line positions in the old version are marked with \textless INSERT\textgreater.
Additionally, \textless BLOCK-SPLIT\textgreater \space functions as a separator between replace blocks.
The rest of the inter-line spaces are labelled with \textless NULL\textgreater.
~\autoref{alg:representation} returns hunk with enriched representation.

% \vspace{-5pt}
\begin{algorithm}[t]
\caption{Edit Representation Translation}
\label{alg:representation}
\scriptsize
\begin{algorithmic}[1]
\STATE \textbf{Input:} An edit hunk labelled in git-diff format, $L$
\STATE \textbf{Output:} An edit hunk labelled in \tool format, $L^*$
\STATE $code\_token\_mapping = LCS(parser(L.old\_version\_code),$
\par\qquad\qquad\qquad\qquad\qquad\qquad\quad\quad$parser(L.new\_version\_code))$ 
\STATE $line\_mapping = token2line\_mapping(code\_token\_mapping)$
\FOR{$block \in line\_mapping$} 
    \STATE $inter\_labels, inline\_labels = convert2label($\par\qquad\qquad$block.old.line\_idx, block.new.line\_idx)$;
    \STATE $L^*.inter\_labels.$extend$(inter\_labels)$;
    \STATE $L^*.inline\_labels.$extend$(inline\_labels)$;
\ENDFOR
\STATE \textbf{ASSERT} $len(L^*.inter\_label) - 1 == len(L^*.inline\_label) == len(L.old\_version\_code\_lines)$
\STATE \textbf{return} $L^*$
\end{algorithmic}
\end{algorithm}
\vspace{-10pt}

%% file: Sections/edit-representation-example-new.tex
\definecolor{codegreen}{rgb}{0.0, 0.66, 0.30}
\definecolor{codered}{rgb}{1,0.05,0.05}
\definecolor{codegray}{rgb}{0.5,0.5,0.5}
\definecolor{codegray}{rgb}{0.5,0.5,0.5}
\definecolor{codeblue}{rgb}{0.0,0.0,0.5}
\definecolor{codered}{rgb}{0.6,0.0,0.0}
\definecolor{codepurple}{rgb}{0.58,0.0,0.82}
\definecolor{codeorange}{rgb}{1.0,0.5,0.0}
\definecolor{codecyan}{rgb}{0.0,0.6,0.6} % New color for strings

\lstdefinestyle{lgeometry}{ xleftmargin = 20pt, xrightmargin = 0pt, frame = tb, framesep = \fboxsep, framexleftmargin = 20pt}

% Define a style for the listings
\lstdefinestyle{gostyle}{
    language=Python,
    basicstyle=\ttfamily\scriptsize,
    breakatwhitespace=false,
    breaklines=true,
    captionpos=b,
    keepspaces=true,
    showspaces=false,
    showstringspaces=false,
    showtabs=false,
    tabsize=2,
    commentstyle=\color{codegray},
    keywordstyle=\color{blue},
    stringstyle=\color{codepurple},
    moredelim=[is][\color{codered}\textbf]{@@}{@@},
    moredelim=[is][\color{codegreen}\textbf]{^}{^},
    columns=flexible,
    aboveskip=-0.4\baselineskip,
    belowskip=-1.3\baselineskip,
    numbers=left,
    numberstyle=\tiny\color{codegray},
    stepnumber=1,
    firstnumber=auto,
    numbersep=4pt,
    % style = lgeometry,
}
\lstset{style=gostyle}

\begin{table*}[t]
\caption{
Enriched edit semantic labelling: The first column denotes inter-line labels, including N (NULL), I (INSERT), and B (BLOCK-SPLIT). The second column denotes inline labels, including K (KEEP), R (REPLACE) and D (DELETE).
}
\label{tab:representation-example}
%\vspace{-5pt}
%\scriptsize
\centering
\small
\tabcolsep=0.06cm
\begin{tabular}{|p{0.01\linewidth}|p{0.01\linewidth}|p{0.01\linewidth}|p{0.51\linewidth}|p{0.01\linewidth}|p{0.40\linewidth}|}

\hline
\multicolumn{4}{|c|}{Enriched edit representation on hunk before edit} & \multicolumn{2}{c|}{Hunk after edit}\\
\hline
  \multirow{1}{*}{N} & & & & & \\
\cline{1-6}
& \multirow{1}{*}{K} & &
\begin{lstlisting}[firstnumber=1]
  def extract_tags(req_data):
\end{lstlisting} & & 
\begin{lstlisting}[firstnumber=1]
  def extract_tags(req_data):
\end{lstlisting}\\
\cline{1-6}
 \multirow{1}{*}{N} & & & & & \\
\cline{1-6}
 & \multirow{1}{*}{K}& &
\begin{lstlisting}[firstnumber=2]
    tags = []
\end{lstlisting} & &
\begin{lstlisting}[firstnumber=2]
    tags = []
\end{lstlisting}\\
\cline{1-6}
  \multirow{1}{*}{N} & & & & & \\
\cline{1-6}
 & \multirow{2}{*}{D}& &
\cellcolor{red!20}
\begin{lstlisting}[firstnumber=3]
-   req_tags = {k: v for k, v in req_data.items() if k.startswith('Tags.member.')}
\end{lstlisting} & & \\
\cline{1-6}
  \multirow{1}{*}{N} & & & \begin{lstlisting}
\end{lstlisting} & & \\
\cline{1-6}
 & \multirow{1}{*}{R}& &
\cellcolor{red!20}
\begin{lstlisting}[firstnumber=4]
-   for i in range(int(len(req_tags.keys()) / 2)):
\end{lstlisting} & & 
\cellcolor{green!20}
\begin{lstlisting}[firstnumber=3]
+   for i in range(1, 200):
\end{lstlisting} \\
\cline{1-6}
 \multirow{2}{*}{I} & & & & & \cellcolor{green!20}\begin{lstlisting}[firstnumber=4]
+     k1='Tags.member.%s.Key'%i
+     k2='Tags.member.%s.Value'%i
\end{lstlisting} \\
\cline{1-6}
 & \multirow{1}{*}{R}& &
\cellcolor{red!20}
\begin{lstlisting}[firstnumber=5]
-     key = req_tags['Tags.member.' + str(i + 1) + '.Key']
\end{lstlisting} & & 
\cellcolor{green!20}
\begin{lstlisting}[firstnumber=6]
+     key = req_data.get(k1)
\end{lstlisting} \\
\cline{1-6}
  \multirow{1}{*}{B} & & & \begin{lstlisting}
\end{lstlisting} & & \\
\cline{1-6}
 & \multirow{1}{*}{R}& &
\cellcolor{red!20}
\begin{lstlisting}[firstnumber=6]
-     value = req_tags['Tags.member.' + str(i + 1) + '.Value']
\end{lstlisting} & &
\cellcolor{green!20}
\begin{lstlisting}[firstnumber=7]
+     value = req_data.get(k2)
\end{lstlisting} \\

\cline{1-6}
 \multirow{2}{*}{I} & & & & & \cellcolor{green!20}\begin{lstlisting}[firstnumber=8]
+     if key is None or value is None:
+       break
\end{lstlisting} \\
\cline{1-6}
 & \multirow{1}{*}{K}& &
\begin{lstlisting}[firstnumber=7]
      tags.append({'Key': key, 'Value': value})
\end{lstlisting} & &
\begin{lstlisting}[firstnumber=10]
      tags.append({'Key': key, 'Value': value})
\end{lstlisting}\\
\cline{1-6}
  \multirow{1}{*}{N} & & & & & \\
\cline{1-6}
 & \multirow{1}{*}{K}& &
\begin{lstlisting}[firstnumber=8]
    return tags
\end{lstlisting} & & 
\begin{lstlisting}[firstnumber=11]
    return tags
\end{lstlisting}\\
\cline{1-6}
  \multirow{1}{*}{N} & & & & &  \\
\hline

\end{tabular}
%\vspace{-15pt}
\end{table*}

%% file: Sections/edit-composition-invoker.tex
\subsection{Edit-composition Invocation}

\subsubsection{Predefined Edit Composition}
We select five predefined edit compositions\footnote{These five composition types reflect basic capabilities widely supported by mainstream LSPs and commonly observed in practice. More advanced compositions depend on specific languages and LSP implementations.},
each is denoted as an edit set $E$,
\begin{itemize}[leftmargin=*]
    \item \textbf{Variable Rename:}  
    Edits in the variable rename composition $E_{var}$ consistently modify all occurrences of the same variable identifier across the codebase.
    \item \textbf{Function Rename}: Similarly, edits of this composition $E_{func}$ replace the same function identifier;
    \item \textbf{Def-Use Propagation:}
    Edits to function signatures (e.g., parameter addition or type change) affect both definitions and usage sites, forming a def\&use composition $E_{defuse}$.
    \item \textbf{Code Clone Update:}
    Edits in code clone composition $E_{cc}$ share similar pre-edit code $code_b$ and post-edit code $code_a$.
    \item \textbf{Code Diagnose Fix:}
    Edits within diagnostic compositions $E_{dia}$ propagate to each other via lint errors. 
\end{itemize}

\subsubsection{Learning Invocation}
\begin{figure}
    \centering
    \includegraphics[width=0.85\linewidth]{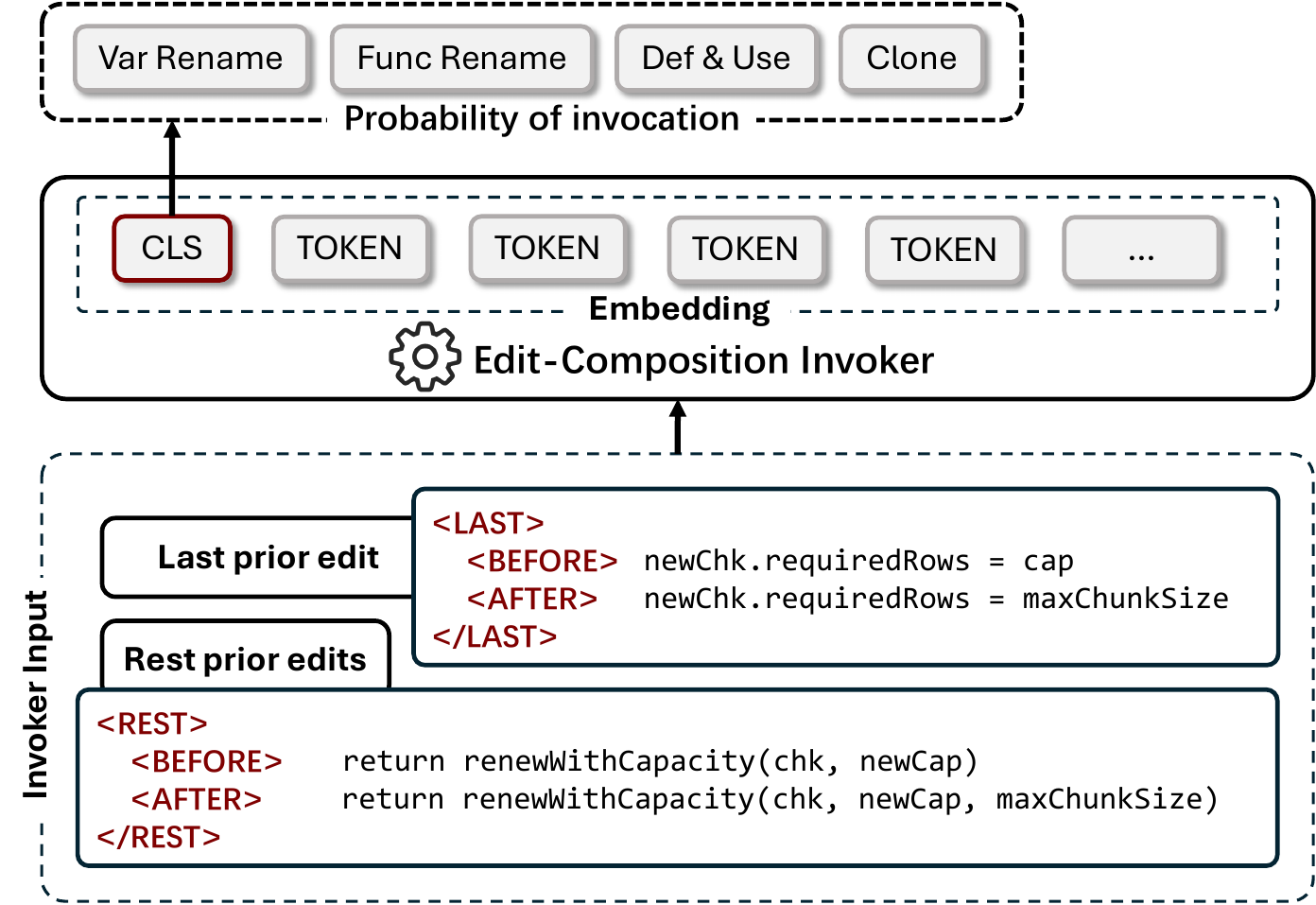}
    %\vspace{-5pt}
    \caption{Overview of edit-composition invoker}
    \label{fig:invoker}
    \vspace{-5pt}
\end{figure}

We design edit composition invoker as a multi-label classifier predicting whether the latest prior edit fits any defined composition type.
\highlight{C3, C9}{We exclude prompt tuning as it is non-trivial for prompts to accurately decide when and which edit composition to invoke across hundreds of potential scenarios.}
Instead, we adopt an encoder model, as shown in \autoref{fig:invoker}:
given prior edits $E_{p}$, 
we divide the edits into the latest edit and the rest of the prior edits, feeding into the encoder. 
The encoder outputs a confidence score for each type of edit composition in CLS position (see \autoref{fig:invoker}), 
where scores above a threshold indicate composition membership.
For each edit, we adopt XML tags including \textless BEFORE\textgreater\ and \textless AFTER\textgreater~as the instructions used for model training.
Here, we exclude the code diagnostics composition from the training dataset, 
as LSP implementations push the diagnosis proactively upon document changes,
without requiring active invocation by the Invoker.
To optimize this prediction, we apply a binary cross-entropy loss over the multi-label output, treating each composition type as an independent label. For a single training instance with predicted logits $\{x_i\}$ for composition $i$, and ground truth labels $\{y_i\}$, the loss is defined as shown in \autoref{eq:bceloss}, here $\sigma(\cdot)$ represents the sigmoid function. 
\begin{equation}\label{eq:bceloss}
\mathcal{L}_{\text{BCE}} = - \sum_{i} \left[ y_i \log \sigma(x_i) + (1 - y_i) \log(1 - \sigma(x_i)) \right]
\end{equation}

% \highlight{\subsubsection{Composition Retrieve} Once the last edit \(e_k\) is identified as part of a predefined composition, the system invokes tool services (e.g., LSP-based rename) to retrieve the remaining edits in the same composition, denoted as \(E \setminus \{e_k\}\). This process is represented as:
% $ e_k \leadsto E \setminus \{e_k\}$.
% Since LSP's rename service can directly apply the edit, providing both the edit location and edit content,
% the retrieved rest of the variable rename composition edits are complete: $E_{var}\setminus \{e_k\}=\{e|e = (f,line_{start},line_{end},code_b,code_a)\}$.
% Other compositions, like $E_{defuse}$, are retrieved via LSP's find definition and reference services, which locate related positions, but the edit labels $op$ and the post-edit code $code_a$ remain unknown and depend on the context. Hence retrieved rest of edits are incomplete: $E\setminus \{e_k\}=\{e|e = (f,line_{start},line_{end},code_b,-)\}$, edit tuple with element missing.
% Hence, those retrieved edits should further be completed by a locator for $op$ and a generator for $code_a$.}

%% file: Sections/locator-and-generator.tex
\subsection{Edit Locator and Edit Generator}

\highlight{C3, C9}{We adopt fine-tuning for both the edit locator and the generator: the encoder-only design localizes edits in a single pass without costly decoding iterations, and fine-tuning allows the enriched edit representation to be exploited more effectively than prompt-based approaches.}

\subsubsection{Edit Locator}

\begin{figure}
    \centering
    \includegraphics[width=0.85\linewidth]{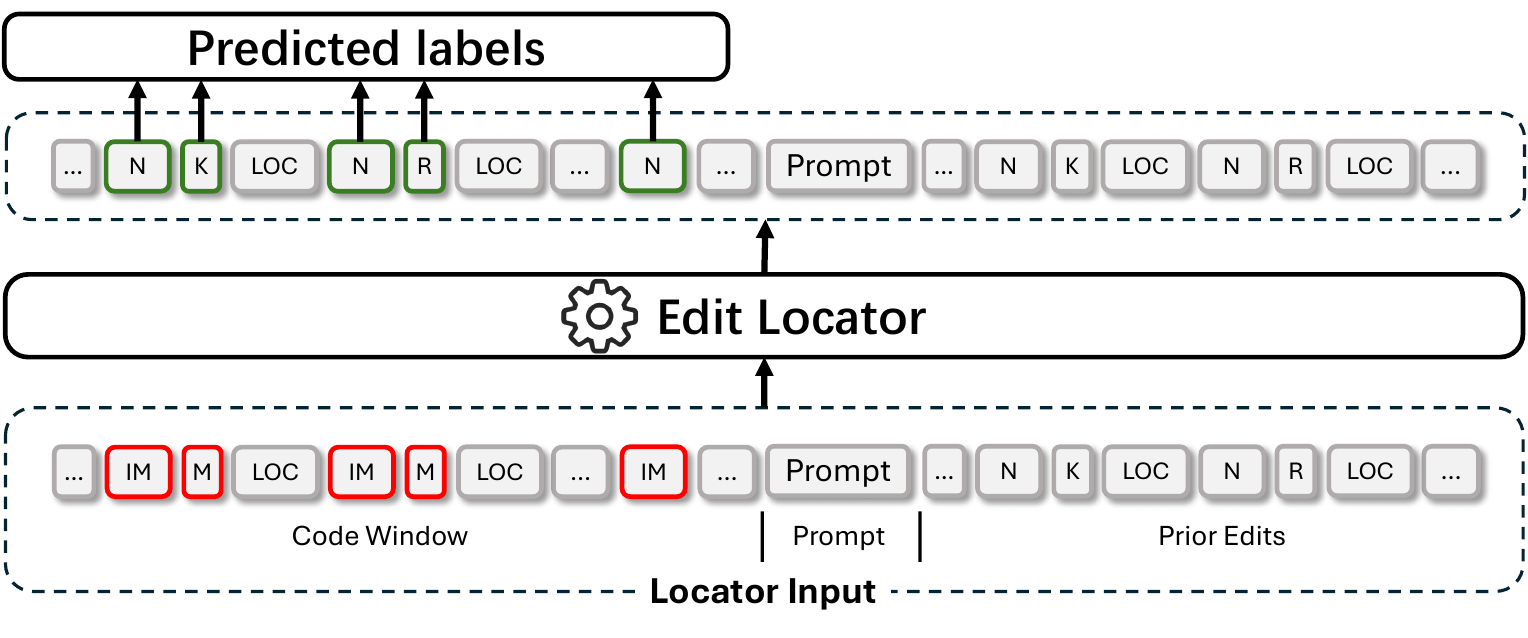}
    %\vspace{-5pt}
    \caption{Neural locator overview: encoder trained to recover the masked tokens. 
    IM is inter-line mask, M is inline mask  
    and LoC denotes a line of code.
    N denotes \textless NULL\textgreater, K denotes \textless KEEP\textgreater, R denotes \textless REPLACE\textgreater. 
    }
    \label{fig:neural-locator}
    \vspace{-5pt}
\end{figure}

Given a code window from $line_{start}$ to $line_{end}$, an optional edit description $prompt$, and selected prior edits $E'_{p}$, the edit locator predicts the edit labels $op$ specified in Section~\ref{sec:edit-representation} by adopting masked language modelling (MLM) \cite{salazar2019masked}.

\autoref{fig:neural-locator} illustrates the details.
For a code window, we place \textless MASK\textgreater \space before each line of code and \textless INTER-MASK\textgreater \space between lines. 
The goal of the neural locator is to recover the two types of mask tokens.
We use cleaned commit messages as $prompt$ during training.
We append selected edits in $E_{p}$ as a way to incorporate prior edits.
Edits are selected as the top-ranked result based on textual similarity, using BM25 \cite{shi2014empirical}.
Edits in $E_{p}$ are formatted in enriched representation with the post-edit code concatenated.

\tool locator is optimized via Cross-Entropy loss on $<$MASK$>$ and $<$INTER-MASK$>$ positions:
\begin{equation}\label{eq:celoss}
    \mathcal{L}_{\text{CE}} = - \sum_{i=1}^{C} y_i \log p_i
\end{equation}
As shown in \autoref{eq:celoss}, C is the total number of classes (e.g., $<$KEEP$>$, $<$REPLACE$>$, $<$DELETE$>$ for $<$MASK$>$, and $<$NULL$>$, $<$INSERT$>$, $<$BLOCK-SPLIT$>$ for $<$INTER-MASK$>$), $y_i$ denotes the one-hot gold label for class i.

\subsubsection{Edit Generator}

\begin{figure}    \centering\includegraphics[width=0.85\linewidth]{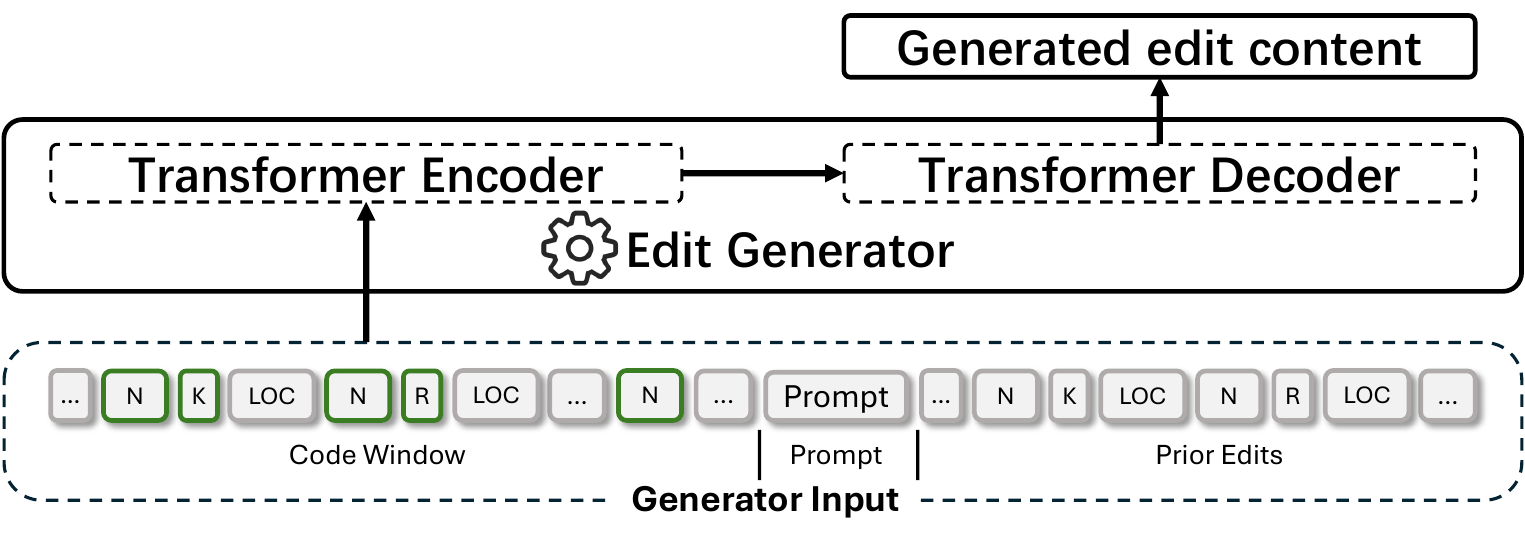}
    %\vspace{-5pt}
    \caption{Overview of neural generator. 
    Given predicted labels like \textless REPLACE\textgreater, 
    a transformer of encoder-decoder generates an edit solution.
    The code window, prompt, and prior edits follow the same format as shown in \autoref{fig:neural-locator}.
    }
    %\vspace{-15pt}
    \label{fig:neural-generator}
\end{figure}

As shown in \autoref{fig:neural-generator}, the generator input includes the code window $code_b$ with edit labels $op$, selected prior edits $E'_{p}$, and optional edit description $prompt$. Each input window only contains one hunk to edit. The generator, formulated as a sequence-to-sequence model \cite{sutskever2014sequence} fine-tuned from an encoder-decoder architecture \cite{vaswani2017attention}, generates \textit{k} candidate edits using beam search \cite{freitag2017beam}.

The generator is fine-tuned using Cross-Entropy loss on the target post-edit code $code_a$, as shown in \autoref{eq:celoss}, where $C$ denotes the vocabulary size.

%% file: Sections/Experiment.tex
\section{Experiment}\label{sec:experiment}

\begin{table}
\centering
\caption{Multilingual benchmark of \tool on 678 repositories and 5 programming languages.}
%\vspace{-5pt}
\label{tab:benchmark}
\begin{tabular}{cccccc} 
\hline
Language                    & Dataset & \#Proj & \#Commit & \begin{tabular}[c]{@{}c@{}}\#Locator \\window\end{tabular} & \begin{tabular}[c]{@{}c@{}}\#Generator \\Hunk\end{tabular}  \\ 
\hline
\multirow{3}{*}{Python}     & Train   & 66     & 5,703    & 72,807                                                     & 36,114                                                      \\
                            & Valid   & 23     & 815      & 10,066                                                     & 5,096                                                       \\
                            & Test    & 36     & 1,630    & 20,026                                                     & 10,078                                                      \\ 
\hline
\multirow{3}{*}{Go}         & Train   & 80     & 4,919    & 66,705                                                     & 32,594                                                      \\
                            & Valid   & 21     & 707      & 9,414                                                      & 4,636                                                       \\
                            & Test    & 64     & 1,405    & 18,457                                                     & 9,206                                                       \\ 
\hline
\multirow{3}{*}{Java}       & Train   & 59     & 9,480    & 135,196                                                    & 67,714                                                      \\
                            & Valid   & 8      & 1,354    & 18,813                                                     & 9,611                                                       \\
                            & Test    & 40     & 2,705    & 38,777                                                     & 19,578                                                      \\ 
\hline
\multirow{3}{*}{Javascript} & Train   & 77     & 1,826    & 21,724                                                     & 11,022                                                      \\
                            & Valid   & 14     & 264      & 3,233                                                      & 1,706                                                       \\
                            & Test    & 33     & 520      & 5,989                                                      & 3,105                                                       \\ 
\hline
\multirow{3}{*}{Typescript} & Train   & 108    & 5,116    & 64,766                                                     & 33,054                                                      \\
                            & Valid   & 26     & 734      & 4,743                                                      & 9,416                                                       \\
                            & Test    & 34     & 1,469    & 18,360                                                     & 9,397                                                       \\ 
\hline
\multirow{3}{*}{Total}      & Train   & 390    & 27,044   & 361,198                                                    & 180,498                                                     \\
                            & Valid   & 92     & 3,874    & 50,942                                                     & 25,792                                                      \\
                            & Test    & 207    & 7,729    & 101,609                                                    & 51,364                                                      \\
\hline
\end{tabular}
\vspace{-10pt}
\end{table}

We evaluate \tool via the following research questions:
\begin{itemize}[noitemsep, topsep=0pt, leftmargin=*]
    \item \textbf{RQ1 (Edit Location, Section~\ref{subsec:rq1})}: Can the edit locator effectively identify editable lines and their types?
    \item \textbf{RQ2 (Edit Generation, Section~\ref{subsec:rq2})}: Can the edit generator effectively generate the edit content?
    \item \textbf{RQ3 (Ablation study, Section~\ref{subsec:rq3})}: Can the enriched edit semantic representation effectively boost the performance of the neural locator and generator?
    \item \textbf{RQ4 (Edit-Composition Invocation, Section~\ref{subsec:rq4})}: Can the edit-composition invoker be triggered accurately?
    \item \textbf{RQ5 (Real-world Simulation, Section~\ref{subsec:rq5})}: What is the performance of \tool in a real-life editing environment?
\end{itemize}

\highlight{C4}{Given the space limit, qualitative analysis and the visualized simulation process are available at \cite{homepage}}.

\textit{\textbf{Experiment Setup: }}
We construct a multilingual benchmark of 5 programming languages (Python, Go, Java, JavaScript, and TypeScript).
We use an open-source LLM (e.g., Llama-3-8B-Instruct \cite{llama3modelcard}, \highlight{C4}{see our anonymous site \cite{homepage} for prompt design}) to filter out commits with vague or multi-intent messages (e.g., adding new features while fixing unrelated bugs), 
as such content hinders learning the semantic link between language and edits. 
\highlight{C21}{Commit messages are typically abbreviated and often contain irrelevant noise such as
PR numbers, repeated messages or committer information, 
leading to a distribution shift from real-life edit descriptions. 
To address this, we refined retained messages for clarity 
(e.g., removing pull request IDs or committer emails). 
Our manual evaluation shows that the refined messages are more natural and concise, and LLaMA3 performs well in this task by effectively removing noise and rewriting messages into fluent descriptions.}
% \highlight{C4}{Other cleaning criteria we adopted including:
% \begin{itemize}[leftmargin=*]
%    \item Commit contains only ASCII characters;
%    \item Commit message length is between 8 and 128 tokens;
%    \item Commit author and committer are not bots;
%    \item Commit is not a merge or pull request commit;
%    \item Commit contains 3-15 hunks across at least 2 files;
%    \item Commit does not contain trivial edits (e.g., deleting empty lines);
%    \item Each edit hunk contains no more than 15 modified lines;
% \end{itemize}
% These criteria enhance data quality while ensuring compliance with language model input length limitations.}

The resulting benchmark comprises 678 top-starred repositories and a final set of 38k commits, \highlight{C4, C15}{statistics are shown in \autoref{tab:benchmark}}. 
Based on the benchmark, we translate the
data format for training different models to predict specific training tasks such as
composition invoking, edit location, and edit generation. 
We adopt cross-project splits, 
with training and test sets from disjoint projects 
to prevent data leakage.

\subsection{RQ1 (Edit location)} \label{subsec:rq1}
\noindent\textit{\textbf{Benchmark:}} To train and evaluate the neural locator, 
we collected 101,609 sliding code windows from the general benchmark, 
each is paired with up to three relevant hunks. 
To avoid data leakage, hunks that overlap with the target code window are excluded.
To better align with real-world deployment scenarios, a single code window may contain multiple edit hunks.
All models, including baselines and \tool, take the same input: a code window, an optional prompt, and selected prior edits. The output is a label sequence.

\noindent \textit{\textbf{Metric:}} We evaluate predicted labels using accuracy, macro-averaged precision, recall, and F1-score.
Macro averaging equally weights all classes, ignoring the class imbalance.

\noindent \textit{\textbf{Baseline:}} \tool locator is compared with the state-of-the-art CoEdPilot locator and naive code clone detector\footnote{The code clone detector is included based on the observation that developers often locate relevant code by searching for similar code snippets.} \cite{codeclonedetector}.
% \footnote{Code lines within code window that match the pre-edit code of prior replace edits are labelled as $<$REPLACE$>$, while the line immediately following a region that matches the prefix context of prior insertion edits is labelled as $<$INSERT$>$.}

All neural models adopt the \texttt{Salesforce/\allowbreak codet5-\allowbreak large} \cite{Salesforcecodet5large, wang2021codet5} encoder, 
fine-tuned on our dataset. 
This experiment precludes LSP, as it requires the simulation of the entire project.
The detailed evaluation for \tool with external tools equipped is in Section~\ref{subsec:rq5}.

\begin{table}
\centering
\caption{RQ1: Edit location performance: \tool uses enriched semantics; baselines use plain representations.}
%\vspace{-5pt}
\label{tab:rq1}
\begin{tabular}{lcccc}
\hline
 & Acc. (\%) & Prec. (\%) & Rec. (\%) & F1 (\%) \\
\hline
\tool Locator & \textbf{93.48} & \textbf{69.61} & \textbf{67.89} & \textbf{68.71} \\
% Plain semantic & 90.78 & 60.76 & \textbf{72.68} & 65.41 \\
CoEdPilot Locator& 83.97 & 48.42 & 61.75 & 51.91 \\
Code Clone Detector& 82.15 & 40.47 & 44.14 & 41.38 \\
\hline
\end{tabular}
\vspace{-5pt}
\end{table}

\noindent \textit{\textbf{Result:}}
\autoref{tab:rq1} shows the locator’s static performance with enriched semantics, 
achieving 69.61\% precision and 68.71\% F1 in locating edit lines.
The strength of our enriched semantic locator lies in high-precision edit identification, 
achieving a 43.76\% improvement over the best baseline.
Meanwhile, the code clone detector performs well mainly for copy-paste edits, but its overall effectiveness is limited.

\subsection{RQ2 (Edit generation)}\label{subsec:rq2}
\noindent\textit{\textbf{Benchmark:}} Similar to the neural locator, we collected 51,364 test hunks from the general benchmark, each paired with 3 other hunks from the same commit.
All generators share the same input, i.e., a code window with edit labels, an optional prompt, and selected prior edits. 
The target output is the post-edit code for the hunk within the window.

\noindent  \textit{\textbf{Metric:}}
We generate 10 candidates per sample, 
ranked by confidence, 
and evaluate performance at Top-1, 3, 5, and 10 using 
exact match rate (EMR) and BLEU-4 \cite{papineni2002bleu}. 
EMR is the percentage of samples with an exact match in the Top-k, 
while BLEU-4 takes the highest score among them.

\noindent \textit{\textbf{Baseline:}} 
We compare our enriched edit semantic generator with 3 state-of-the-art edit generators: 
CoEdPilot \cite{code-edit-pilot}, GrACE \cite{grace} and CCT5 \cite{lin2023cct5}.
All models adopt \texttt{Salesforce/\allowbreak codet5-base} as the base model and are
fine-tuned on our dataset. 

\begin{table}
\centering
\caption{RQ2: static generation performance: \tool uses enriched semantics; baselines use plain representations.}
\label{tab:rq2}
%\vspace{-5pt}
\begin{tabular}{lccccc}
\hline
\textbf{Model} & \textbf{Metric} & \textbf{@1} & \textbf{@3} & \textbf{@5} & \textbf{@10} \\
\hline
\multirow{2}{*}{\tool generator} & \textbf{EMR} & \textbf{48.02} & \textbf{53.44} & \textbf{55.04} & \textbf{57.05} \\
 & \textbf{BLEU} & \textbf{69.68} & \textbf{73.55} & \textbf{74.73} & \textbf{76.10} \\
\hline
% \multirow{2}{*}{Plain semantic} & \textbf{EMR} & 44.71 & 50.08 & 51.70 & 53.61 \\
%  & \textbf{BLEU} & 65.82 & 70.14 & 71.41 & 72.81 \\
% \hline
\multirow{2}{*}{CoEdPilot generator} & \textbf{EMR} & 43.20 & 48.92 & 50.61 & 52.70 \\
 & \textbf{BLEU} & 64.90 & 69.64 & 70.94 & 72.48 \\
\hline
\multirow{2}{*}{GrACE} & \textbf{EMR} & \multicolumn{1}{r}{39.54} & \multicolumn{1}{r}{46.30} & \multicolumn{1}{r}{48.45} & \multicolumn{1}{r}{50.88} \\
 & \textbf{BLEU} & \multicolumn{1}{r}{61.89} & \multicolumn{1}{r}{68.47} & \multicolumn{1}{r}{70.33} & \multicolumn{1}{r}{72.36} \\
\hline
\multirow{2}{*}{CCT5} & \textbf{EMR} & 39.76 & 45.76 & 47.77 & 51.45 \\
 & \textbf{BLEU} & 62.91 & 67.88 & 69.43 & 72.43 \\
\hline
\end{tabular}
\vspace{-5pt}
\end{table}

\noindent \textit{\textbf{Result:}}
\autoref{tab:rq2} shows the static generation performance of edit generation with Top-10 candidates.
Our generator with enriched edit semantics achieves the best performance in generating edit options: 
the Top-1 candidate has an exact match rate of 48.02\%, 
yielding a notable boost of 11.16\% compared with the state-of-the-art generator models.
In the Top-10 options, an exact match can be found in more than 57\% of the cases,
suggesting potential gains in user efficiency.

% It's noteworthy that bringing the enriched semantics results in a 7.40\% improvement compared with plain semantics, largely because the enriched edit semantics provided clearer edit instructions for target code windows.

\subsection{RQ3 (Ablation study)} \label{subsec:rq3}
\noindent \textit{\textbf{Benchmark:}} 
We evaluate the edit locator and generator using benchmarks from Sections~\ref{subsec:rq1} and~\ref{subsec:rq2}.

\noindent \textit{\textbf{Metric:}}
Same metrics as in Section~\ref{subsec:rq1} and Section~\ref{subsec:rq2}.

\noindent \textit{\textbf{Baseline:}} 
We compare our enriched edit semantic representation with the 3-label one by fine-tuning models of the same size on the same dataset, differing only in representation.

\noindent \textit{\textbf{Result:}}
As shown in \autoref{tab:rq3}, our enriched edit semantics achieve 69.61\% precision, a 14.57\% improvement over the plain representation.
We observed that the plain semantics locator achieves slightly higher recall, 
likely due to having fewer labels, 
which simplifies classification with coarser decision boundaries and fewer misclassifications. 
However, for edit location recommendation, we prioritize precision over recall, 
as developers benefit more from accurate suggestions than from broader but less precise coverage.
It is noteworthy that enriched semantics improve generator performance by 7.40\% over plain semantics,
mainly by offering clearer edit instructions for target code windows.

\begin{table}
\tabcolsep=0.07cm
\centering
\footnotesize
\caption{RQ3: Ablation study.}
%\vspace{-5pt}
\label{tab:rq3}
\begin{tabular}{ccccc|ccccc} 
\hline
\multirow{3}{*}{Model}    & \multicolumn{4}{c|}{Locator}                                                                                                                                                                                     & \multicolumn{5}{c}{Generator}                                                             \\ 
\cline{2-10}
                          & \begin{tabular}[c]{@{}c@{}}Acc.\\(\%)\end{tabular} & \begin{tabular}[c]{@{}c@{}}Prec.\\(\%)\end{tabular} & \begin{tabular}[c]{@{}c@{}}Rec.\\(\%)\end{tabular} & \begin{tabular}[c]{@{}c@{}}F1\\(\%)\end{tabular} & \multicolumn{1}{l}{} & @1             & @3             & @5             & @10             \\ 
\hline
\multirow{2}{*}{Enriched} & \multirow{2}{*}{\textbf{93.48}}                    & \multirow{2}{*}{\textbf{69.61}}                     & \multirow{2}{*}{67.89}                             & \multirow{2}{*}{\textbf{68.71}}                  & EMR                  & \textbf{48.02} & \textbf{53.44} & \textbf{55.04} & \textbf{57.05}  \\
                          &                                                    &                                                     &                                                    &                                                  & BLEU                 & \textbf{69.68} & \textbf{73.55} & \textbf{74.73} & \textbf{76.10}  \\ 
\hline
\multirow{2}{*}{Plain}    & \multirow{2}{*}{90.78}                             & \multirow{2}{*}{60.76}                              & \multirow{2}{*}{\textbf{72.68}}                    & \multirow{2}{*}{65.41}                           & EMR                  & 44.71          & 50.08          & 51.70          & 53.61           \\
                          &                                                    &                                                     &                                                    &                                                  & BLEU                 & 65.82          & 70.14          & 71.41          & 72.81           \\
\hline
\end{tabular}
\vspace{-5pt}
\end{table}

\subsection{RQ4 (Edit-composition invocation)} \label{subsec:rq4}

In this experiment, we leverage the facilities of LSPs \cite{pyright, gopls, jdtlsp, jstslsp} as the IDE edit composition.

\noindent \textit{\textbf{Benchmark:}}\label{sec:invoker_benchmark}
\highlight{C4, C18}{Given a commit of $N$ edit hunks, we randomly select one target hunk $H_t$ and up to 2 background hunks $H_{b1}$, $H_{b2}$. 
We configure the project state such that $H_t$, $H_{b1}$, $H_{b2}$ are applied (post-edited) while the remaining $N-3$ hunks remain unapplied (pre-edited), simulating a partial commit completion scenario.
We invoke LSP services (rename, find reference, find clone) at $H_t$'s location. If the service returns edit locations matching any of the remaining $N-3$ unedited hunks (determined by line index overlap), we label this as a positive sample with $H_t$, $H_{b1}$, $H_{b2}$ as input and the corresponding LSP service type as output. The dataset contains 8,534 training, 1,294 validation, and 2,499 test samples.}

\noindent \textit{\textbf{Metrics:}}
We use macro-averaged precision, recall, and F1-score 
to assess the edit composition invocation performance.

\noindent \textit{\textbf{Baseline:}}
We compare our method against two baselines: blindly invoking all LSP services, and randomly invoking one.

\begin{table}
\centering
\tabcolsep=0.07cm
\caption{RQ4: The performance of Invoker.}
\label{tab:rq4}
%\vspace{-5pt}
\begin{tabular}{clccc} 
\hline
\multicolumn{2}{l}{} & Precision (\%) & Reccall (\%) & F1 (\%) \\
\hline
\multirow{5}{*}{\begin{tabular}[c]{@{}c@{}}Edit-Composition \\Invoker\end{tabular}} & Variable Rename & 91.15 & 98.59 & 94.72 \\
 & Function Rename & 98.71 & 97.44 & 98.07 \\
 & Def \& use & 84.06 & 87.90 & 85.94 \\
 & Clone & 95.88 & 94.59 & 95.23 \\ 
\cline{2-5}
 & Average & \textbf{92.45} & 94.63 & \textbf{93.49} \\ 
\hline
\multicolumn{2}{c}{Blindly invoking} & 22.01 & \textbf{100.00} & 35.50 \\
\hline
\multicolumn{2}{c}{Randomly invoking} & 21.85 & 50.26 & 29.71 \\
\hline
\end{tabular}
\vspace{-10pt}
\end{table}

\noindent \textit{\textbf{Result:}}
\autoref{tab:rq4} shows the performance of the edit-composition invoker, which achieves an F1 score of 93.49\%, 
substantially outperforming both blindly and randomly invoking LSP functions. 
The results indicate that 
the edit composition can be effectively invoked.

\subsection{RQ5 (Real-world simulation)} \label{subsec:rq5}
\noindent \textit{\textbf{Benchmark:}} 
500 commits (100 for each language) are randomly selected from the test set in the general benchmark, 
comprising a total of 3,211 edit hunks.

\noindent \textit{\textbf{Simulation process:}}
Compared with the previous two research questions (Section~\ref{subsec:rq1} and Section~\ref{subsec:rq2}),
real-world editing scenarios are much more challenging, because:
(1) the majority of the code remains unchanged
(2) with fewer prior edits available; and
(3) the impact of a single edit is limited—it rarely propagates to all edits within a commit and, in some cases, does not propagate at all.
Hence, to comprehensively evaluate the actual performance in a real-world editing scenario,
we simulate the process of a programmer making a commit by editing each hunk until the old version is transformed into the new version.
\highlight{C4, C19}{
We define the simulation process as 4 stages.
1) \textbf{Initialization}: For a given commit, we check out the project to its pre-commit version and use git diff to identify all edit hunks. The first hunk listed by git diff is designated as the initial edit, which is applied to obtain the starting project state.
2) \textbf{Location prediction and selection}: 
The process then iterates over the remaining edits. 
At each step, the locator model predicts candidate locations for the next edit based on the commit message and prior edits. Candidates are ranked by confidence scores, either produced by the model or set to a default value of 1.0 for locations obtained via LSP services. 
A predicted location is considered a match to a ground-truth edit if it has more than 50\% line overlap with the ground-truth hunk.
If a match is found, it is passed to the generator; otherwise, the virtual programmer randomly selects one remaining ground-truth location.
3) \textbf{Edit generation and application}: Given the selected location, 
the generator produces 10 candidate edits, which are compared with the ground truth. 
The virtual programmer applies the ground-truth edit content at the selected location to the project before proceeding. 
This quality control mechanism prevents the simulation from continuing from erroneous states, 
emulating human oversight in interactive editing.
4) \textbf{Iteration}: Steps 2 and 3 repeat until all edits in the commit are simulated and applied.}
A video of this simulation process is available at \cite{homepage}.

\noindent \textit{\textbf{Metric:}}
For localization, we report the \textbf{Top-K match rate (MR@K)} and the \textbf{time cost}.
MR@K denotes the percentage of predictions with at least one correct location in the Top-K. 
Time cost reflects the average latency of location prediction. 
\highlight{C3, C17}{Both metrics are affected by Invoker and Locator performance; component failures lead to lower MR@K and higher time costs.}
For generation, we analyze the \textbf{BLEU-4 score distribution} in three bands: 
100 (directly usable for user), 50–100 (minor edits needed), and $<$50 (likely rejected). 
\highlight{C3, C17}{BLEU-4 scores are affected by both Invoker and Generator; component failures lead to score drops.}
\textbf{Acceptance rate @K} denotes the percentage of Top-K suggestions containing at least one match with the gold location and a BLEU-4 of 100, \highlight{C3, C17}{which is affected by all three components}.

\noindent \textit{\textbf{Baselines:}} \tool composes a locator and a generator with enriched edit semantics, integrated with LSP \cite{pyright, gopls, jdtlsp, jstslsp} and Invoker. 
We compare \tool to baselines:
\begin{enumerate}[leftmargin=*]
    \item \textbf{\tool\xspace w/o Invoker}: \tool without Invoker, blindly invoking LSP services;
    \item \textbf{Enriched semantic}: Neural locator and generator, both with the enriched edit semantics (6-label representation);
    \item \textbf{Plain semantic}: Neural locator and generator, both of plain edit semantics (3-label representation);
    \item \textbf{CoEdPilot}: Locator and generator from \coedpilot, models re-trained on the same dataset and architecture as \tool;
    \item \textbf{CCD}: Code clone detector as locator and neural generator of plain edit semantics.
    \item \highlight{C3, C14}{\textbf{Cursor}: 
Since Cursor of version 0.46 lacks APIs for large-scale simulation, 
to simulate a commit, each time we manually apply an edit to the project. 
If this edit triggers Cursor's Tab recommendation, we follow the single suggested edit and mark it as Top-1.
Otherwise, we input the prior edit and edit description into the Cursor Chat, which provides multiple edit suggestions.
We rank all suggestions by their proximity to the last applied edit and evaluate whether the ground truth appears in the Top-K ranked recommendations.}
\end{enumerate}

\begin{table}
\centering
\tabcolsep=0.05cm
\caption{RQ5: Real-world edit simulation performance.
% : \tool refers to our proposed framework, \tool \xspace- Invoker refers to our framework without Invoker. Enriched semantic refers to the \tool without LSP and Invoker. Plain semantic refers to the combination of 3-label locator and generator. CCD refers to a code clone detector as the locator with 3-label generator.
}
\label{tab:rq5}
%\vspace{-5pt}
\begin{tabular}{lcccc|ccc} 
\hline
\multicolumn{1}{c}{\multirow{4}{*}{\textbf{Model}}}  & \multicolumn{4}{c|}{\textbf{Locating}}                                                                                                                                                  & \multicolumn{3}{c}{\textbf{Generation}}                                                                   \\ 
\cline{2-8}
\multicolumn{1}{c}{}                                 & \multicolumn{3}{c}{\begin{tabular}[c]{@{}c@{}}\textbf{Match}\\ \textbf{rate (\%)}\end{tabular}} & \multirow{2}{*}{\begin{tabular}[c]{@{}c@{}}\textbf{Time }\\\textbf{(s)}\end{tabular}} & \multicolumn{3}{c}{\begin{tabular}[c]{@{}c@{}}\textbf{BLEU-4}\\ \textbf{distribution (\%)}\end{tabular}}  \\ 
\cline{2-4}\cline{6-8}
\multicolumn{1}{c}{}                                 & \textbf{@1} & \textbf{@3} & \textbf{@5}                                                         &                                                                                       & \textbf{100} & \textbf{50\textasciitilde{}100} & \textbf{$<$ 50}                     \\ 
\hline
\tool                                 & \textbf{35.18}       & \textbf{42.07}       & \textbf{44.24}                                                              & 3.27                                                                                  & \textbf{49.23}        & 24.32                           & 26.45                                                    \\ 
\hline
\tool\xspace w/o Invoker & 32.52       & 39.80       & 42.23                                                               & 3.47                                                                                  & \textbf{49.23}        & \textbf{24.52}                           & \textbf{26.25}                                                    \\ 
\hline
Enriched semantic                                    & 32.56       & 40.71       & 43.39                                                               & 3.82                                                                                  & \textbf{49.23}       & \textbf{24.52}                        & \textbf{26.25}                                                    \\ 
\hline
Plain semantic                                       & 31.05       & 36.35       & 37.57                                                               & 3.81                                                                                  & 45.41        & 23.67                           & 30.92                                                    \\ 
\hline
CoEdPilot                                            & 11.47       & 24.41       & 29.25                                                               & 3.80                                                                                  & 43.17        & 23.81                           & 33.02                                                    \\ 
\hline
CCD                                                  & 15.99       & 15.99       & 15.99                                                               & \textbf{2e-4}                                                                                  & 45.35        & 23.62                           & 31.03                                                    \\
\hline
\end{tabular}
\vspace{-10pt}
\end{table}

\noindent \textit{\textbf{Result:}}
\autoref{tab:rq5} shows the performance in the simulation.
Our \tool outperforms all baselines in locating performance, with over 35\% of cases yielding a useful recommendation at the Top-1 location.
\highlight{C3, C7}{Compared to the state-of-the-art CoEdPilot, \tool achieves a notable 206.71\% improvement in MR@1.}
Integrating LSP and Invoker improves MK@1 performance by over 8.05\% over pure neural solutions,
and cuts locating time by 14.40\%.
Meanwhile, blindly invoking the LSP service is prone to introducing false positive edits, 
reducing performance to that of neural-only solutions.
\highlight{C3, C7}{The enriched semantic approach shows a 15.49\% improvement in MR@5 over plain semantic and a 47.83\% improvement over CoEdPilot, confirming the effectiveness of our enriched edit representation as an additional component alongside Edit-composition Invoker and LSP.}
Note that the code clone detector has the same MR@K, 
as it cannot rank results and marks all as Top-1.
Although code clone detection excels in time efficiency, its recommendation quality remains insufficient for practical use.
We also observed a time discrepancy between \coedpilot and its original work. 
The higher time cost arises as the locator scans more sliding windows ($n$) per query and selects from $m$ prior edits per window via a neural network, adding $O(mn)$ time complexity.

For the generator, our enriched semantic model generates helpful edit solutions (BLEU-4$>$50) in over 73\% of the cases, 
with around 50\% of them achieving BLEU-4=100. 
Compared to the state-of-the-art \coedpilot, \tool achieves a performance boost of 14.04\% in terms of high-quality suggestion (BLEU-4=100) and a 19.90\% drop in low-quality suggestion (BLEU-4$<$50).
\highlight{C3, C7}{Compared to plain semantic and the state-of-the-art \coedpilot, \tool achieves 8.41\% and 14.04\% improvements respectively in high-quality suggestions (BLEU-4=100), while reducing low-quality suggestions (BLEU-4$<$50) by 16.89\% and 19.90\% respectively,
confirming the effectiveness of the TRACE generator and enriched edit representation.}

\autoref{tab:acceptance_rate} shows the acceptance rate of baselines, where \tool is on par with Cursor while achieving a 6\% improvement.
Cursor’s \texttt{Tab} feature offers fast and accurate suggestions but triggers conservatively, 
while its Chat interface struggles to locate cross-file edits proactively. 
As a result, only 8.82\% of accepted edits are cross-file. 
In contrast, \tool leverages LSP integration to support efficient cross-file localization, with 38.46\% of accepted edits being cross-file edits.
% this percentage is calculated on the 30 python commits.
% \usepackage{multirow}

\begin{table}[]
\centering
\caption{RQ5: Acceptance rate in real-world simulation.}
%\vspace{-5pt}
\label{tab:acceptance_rate}
\begin{tabular}{l|ccc}
\hline
\multicolumn{1}{c|}{\multirow{2}{*}{\textbf{Model}}}                & \multicolumn{3}{c}{\textbf{Acceptance rate (\%)}} \\
                \cline{2-4}
                & \textbf{@1}          & \textbf{@3}          & \textbf{@5}          \\
\hline
\tool           &\textbf{25.71}&\textbf{28.54}&\textbf{29.55}\\
\tool\xspace w/o Invoker &24.70&27.73&28.74\\
Enriched        &24.49&27.73&28.95\\
Plain           &21.46&25.91&26.52\\
CoEdPilot       &8.30&17.81&20.65\\
CCD             &12.96&12.96&12.96\\
Cursor          &  24.22      & 26.19       &   27.20      \\
\hline
\end{tabular}
\vspace{-10pt}
\end{table}

%% file: Sections/user_study.tex
\section{User Study}
To validate \tool in real-world use, 
we implement \tool as a VS Code extension and design this user study.
\highlight{C4}{Video recordings and user interaction data are provided at our homepage \cite{homepage}.}

\noindent \textit{\textbf{Baseline:}}
We compare \tool with:
(1) CoEdPilot as the state-of-the-art subsequent edit suggestion system, and
(2) Cursor \cite{cursor} (version 0.46), a popular AI-powered IDE.
During evaluation, we impose no restrictions on Cursor's functionality, and all tools can access IDE LSP services to ensure fairness.

\noindent \textit{\textbf{Participant:}}
\highlight{C5, C20}{We recruit 24 Computer Science students from three universities in both China and Singapore}.
All participants completed a pre-study questionnaire on their experience with programming and AI-assisted tools.
See our website for more demographic details \cite{homepage}.
Participants are stratified into three balanced groups based on their self-reported programming proficiency and AI tool experience, to ensure comparable skill levels across groups. 
The experimental group (EG) uses \tool, while control group 1 (CG1) uses \coedpilot and control group 2 (CG2) uses Cursor.

\noindent\textit{\textbf{Editing task:}}
Participants are asked to reproduce edits from 3 real-world GitHub commits, 
each under a 30-minute time budget.
Selected tasks represent common editing scenarios, with all necessary domain knowledge provided, ensuring completion by participants with general programming experience.
Performance is measured by task completion and time.
\begin{itemize}[noitemsep, topsep=0pt, leftmargin=*]
    \item \textbf{Task 1}: \highlight{C5, C20}{Refactor the \texttt{if} condition from \texttt{if "http" in XX} to \texttt{if XX.startswith("http")} to improve the robustness of string matching \cite{AUTOMATIC1111_userstudy}}, requiring 8 edits across 5 files;
    \item \textbf{Task 2}: \highlight{C5, C20}{Add a \texttt{train\_data\_size} flag to the BERT data pipeline, allowing users to limit training samples and triggering changes along the call chain \cite{models}, requiring 9 edits across 2 files along the call chain};
    \item \textbf{Task 3}: \highlight{C5, C20}{Add the \texttt{noise\_shape} and \texttt{seed} arguments to the Dropout layer API in Keras, enhancing control over the dropout mask shape and randomness seed \cite{keras}}, requiring 5 edits in a single file.
\end{itemize}

\noindent\textit{\textbf{Setup:}}
Participants first complete a warm-up tutorial to familiarize themselves with the assigned tool. 
For each task, they receive 
1) the project code, 
2) background knowledge like code functionality and API usage,
3) a detailed edit description,  
4) the first edit as a hint, and 
5) test cases for validating edits via execution.
\highlight{C5, C20}{The initial edit for each task is selected to provide context and guide subsequent edits: for Task 1, any edit can be randomly chosen due to the uniform editing pattern; for Task 2, the call chain entry point; and for Task 3, the documentation for new arguments.}.
Screens are recorded for analysis. 

\begin{table}
\centering
\tabcolsep=0.05cm
\caption{Performance of EG (\tool), CG1 (CoEdPilot) and CG2 (Cursor), the time cost in minutes.}
%\vspace{-5pt}
\label{tab:user_study}
\begin{tabular}{cccc|cccc|cccc}
\hline
\textbf{EG}            & \textbf{T1} & \textbf{T2} & \textbf{T3} & \textbf{CG1}           & \textbf{T1} & \textbf{T2} & \textbf{T3} & \textbf{CG2}           & \textbf{T1} & \textbf{T2} & \textbf{T3}  \\
\hline
\textbf{P1}   & 3.32        & 7.18        & 8.78        & \textbf{P9}   & 3.55        & 8.42        & 15.60       & \textbf{P17}  & 15.68       & 11.25       & 4.87        \\
\textbf{P2}   & 5.22        & 5.78        & 5.15        & \textbf{P10}  & 3.58        & 13.03       & 10.03       & \textbf{P18}  & 15.87       & 19.28       & 3.42         \\
\textbf{P3}   & 3.83        & 13.37       & 3.50        & \textbf{P11}  & 6.62        & 13.12       & 9.60        & \textbf{P19}  & 17.10       & 18.20       & 4.83         \\
\textbf{P4}   & 3.13        & 8.77        & 4.47        & \textbf{P12}  & 7.90        & 13.07       & 6.70        & \textbf{P20}  & 17.82       & 14.55       & 3.05         \\
\textbf{P5}   & 3.52        & 11.18       & 4.68        & \textbf{P13}  & 4.25        & 12.55       & 7.83        & \textbf{P21}  & 17.08       & 20.12       & 2.98         \\
\textbf{P6}   & 2.92        & 11.00       & 5.62        & \textbf{P14}  & 3.92        & 11.72       & 5.52        & \textbf{P22}  & 13.68       & 30.00       & 2.82         \\
\textbf{P7}   & 4.73        & 8.52        & 3.67        & \textbf{P15}  & 5.18        & 16.38       & 4.57        & \textbf{P23}  & 16.98       & 12.03       & 2.93         \\
\textbf{P8}   & 2.97        & 11.37       & 3.20        & \textbf{P16}  & 8.65        & 18.08       & 18.32       & \textbf{P24}  & 14.63       & 7.30        & 3.10         \\
\hline
\textbf{Avg.} & 3.70        & 9.65        & 4.88        & \textbf{Avg.} & 5.46        & 13.30       & 9.77        & \textbf{Avg.} & 16.11       & 16.59       & 3.50         \\
\hline
\end{tabular}
\vspace{-10pt}
\end{table}

\noindent \textit{\textbf{Metric:}}
We evaluate user study via three metrics:
\textbf{average time cost} for user efficiency,
\textbf{Wilcoxon Signed Rank Test p-values} for statistical significance between groups ($p$ $<$ 0.05 denotes significance),
and \textbf{Effect Size} ($r$) to quantify the difference magnitude ($r \geq 0.5$ denotes a large effect).

\noindent\textit{\textbf{Result:}}
Participants' performance in completing 3 editing tasks is shown in \autoref{tab:user_study}, with the following observations:
\begin{enumerate}[leftmargin=*]
    \item \textbf{Task 1}: EG outperforms both CG1 ($p = 0.0781, r = 0.62$) and CG2 ($p=0.0078,r=0.94$, statistically significant);
    \item \textbf{Task 2}: EG significantly outperforms both CG1 ($p=0.0156,r=0.85$) and CG2 ($p=0.0234,r=0.80$);
    \item \textbf{Task 3}: CG2 significantly outperforms both EG ($p=0.0390,r=0.72$) and CG1 ($p=0.0078,r=0.94$).
\end{enumerate}

\highlight{C5, C20}{
To analyze user performance, 
we first instrumented the extension to monitor user actions in real-time, 
tracking LSP trigger events from user edits and recording user responses to recommendations (accept, reject, or modify). 
Second, we manually analyzed video recordings with three researchers, 
focusing on sessions with notably fast/slow task completion times, plus randomly selected sessions with typical completion times. 
Analysis examined user responses to edit recommendations and behaviors during test failures, 
with cross-validation among researchers.}
Based on this analysis, we provide the following explanations:

\noindent \textbf{Why do \tool and \coedpilot outperform Cursor in Task 1?} Both \tool (EG) and \coedpilot (CG1) support project-wide edit localization across files.
\tool further leverages LSP-based clone detection for efficient cross-file propagation.
In contrast, Cursor (CG2) relies on users to specify files,
which is time-consuming when relevant files are unknown to users.
Additionally, 
Multi-file rewriting in Cursor is slow and disruptive to mental flow.

\noindent \textbf{Why does \tool outperform both \coedpilot and Cursor in Task 2?}
Task 2 involves 9 edits across 2 files,
which are distant but syntactically coherent.
Still taking advantage of tool deduction,
\tool can identify the subsequent edit location via the LSP service,
avoiding the exhaustive file scanning, as in \coedpilot.
In contrast, Cursor users struggle due to its reliance on full-file rewriting and limited cross-file localization ability. 
For example, user P22 manually searched across files but ended up in the wrong one and exhausted the time budget.
%until exhausting the time limit and still failed to pass test cases.

\noindent \textbf{Why does Cursor outperform \tool and \coedpilot in Task 3?}
Given the shared context among 5 edits,
Cursor allows users to quickly trigger the next edit recommendation via \texttt{Tab},
or generate all correct edits in a single rewrite via Chat,
which significantly improves editing efficiency.
Despite lacking edit compositions in this task,
\tool still outperforms \coedpilot by fewer false positive suggestions 
with its improved predicting performance.
%and superior semantic understanding capability.

\noindent \textbf{How do users respond to false positives (over-trust phenomenon)?}
Video analysis reveals an \textit{over-trust} phenomenon in nearly all users, regardless of their AI tool or programming experience. During warm-up, over 92\% of participants quickly gained confidence after a few correct predictions, leading to less caution in accepting recommendations. When test cases failed, users took considerable time to revisit prior decisions, damaging performance. 
We believe that improving human-computer interaction (e.g., intuitive undo/review) and explainable AI (e.g., rationale behind suggestions) is more crucial than further boosting model accuracy, and we plan to explore this direction.

%% file: Sections/Related_work.tex
\section{Related Work}
\textbf{Edit Localization} identifies editable regions based on prior edits, yet most works focus on fault localization. Spectrum-based (SBFL) \cite{abreu2007accuracy, zhang2011localizing, wong2007effective} and mutation-based (MBFL) \cite{moon2014ask, papadakis2015metallaxis} methods localize faults via test outcomes or code mutations. Neural models like Toggle \cite{hossain2024deep} and LLMAO \cite{yang2024large} apply attention or transformers to rank suspicious code. General edit localization is more challenging and underexplored, often relying on static tools. LASE \cite{jacobellis2013lase} uses heuristics to extract edit patterns; CCDemon \cite{lin2015clone} leverages clone detection. However, these methods capture limited propagation types with low precision and provide no actionable instructions.

\textbf{Code Edit Generation:} LLMs are widely applied in software engineering for code \cite{ni2023lever, cai2025automated}, comment \cite{cai2024fly} and test generation \cite{lemieux2023codamosa, qi2025intention, ren2023api,lin2021graph, lin2020recovering, liu2025guipilot}. Among these, code edit generation is especially demanded \cite{alaboudi2021edit}, aiming to suggest edits for a given snippet. Codit \cite{CODIT} proposed a tree-based model; Recoder \cite{recoder} fused code, AST, and paths. CURE \cite{jiang2021cure}, CoditT5 \cite{ZhangETAL22CoditT5}, and CCT5 \cite{lin2023cct5} introduced edit-oriented pre-training. Overwatch \cite{overwatch} modeled temporal edit sequences, while GrACE \cite{grace} encodes both target and prior edits in prompts. Our model further leverages enriched edit semantics for finer control and efficient learning from prior edits.

\highlight{C3, C11, C16}{\textbf{Tool Invocation:} Recent advances have integrated static analysis tools for coding tasks. 
Pei \textit{et al.} \cite{pei2023better} adopted LSP to retrieve function signatures for function call infilling. 
ContextModule \cite{guan2024contextmodule} and Blinn \textit{et al.} \cite{blinn2024statically} extended this approach for more general code completion task.
Recent studies \cite{blyth2025static, li2024iris, li2024enhancing} also combine LLMs with static analysis to improve code quality, including fixing vulnerabilities and readability issues.
% : Blyth \textit{et al.} \cite{blyth2025static} iteratively guide LLMs to fix security and readability issues, IRIS \cite{li2024iris} detects whole-repository vulnerabilities beyond traditional tools, and LLift \cite{li2024enhancing} finds Use Before Initialization bugs in the Linux kernel that static analysis alone misses. 
However, these approaches do not directly support interactive code editing or predict subsequent edit locations.
For editing tasks, 
CodePlan \cite{bairi2024codeplan} employs dependency graphs and LLM-based planning for repository-level tasks, 
while MarsCode Agent \cite{liu2024marscode} introduces multi-agent collaboration with LSP services and code knowledge graphs for bug fixing.
These works differ from TRACE in several key aspects: they are all agentic solutions focused on solving SWE-bench-like tasks \cite{jimenez2023swe}, where, given a project, editing requirements, and tests, they automatically complete edits to make the modified project pass the tests. Compared to TRACE's task, this type of task has lower latency requirements. Additionally, although they all utilize static tools, they serve merely as retrieval mechanisms to provide additional context rather than for edit localization. Moreover, MarsCode Agent's use of LSP functionality is relatively limited.
}

%% file: Sections/Conclusion.tex
\section{Conclusions}
This paper introduces \tool, a subsequent code editing solution that effectively captures the coherence of project-wide code edits. 
\tool proposes Invoker, which integrates LSP to capture edit composition and introduces enriched edit semantics for more accurate representation.
In our experiment, \tool significantly improves edit localization and generation while demonstrating high performance in interactive editing settings, establishing itself as a new state-of-the-art solution for the end-to-end code editing task.
% Beyond that, our user study also reveals users’ over-reliance on AI recommendations, highlighting the need for improved UI design and interpretability in future work.